\documentclass[trackchanges]{aastex701}
\usepackage{amsmath} 
\usepackage{booktabs}
\usepackage{tablefootnote}
\usepackage{multirow}
\usepackage{makecell}

\newcommand{\cellstack}[2][c]{%
  \raisebox{-\height}{\shortstack[#1]{#2}}%
}
\begin{document}
\newcommand*{\hy}{\color{blue}}

\title{Forward Modeling of Spectral-Line Polarization for Weak Magnetic-Field Imaging}

\author[0000-0001-7205-2449]{Chuanpeng Hou}
\affiliation{Institut für Physik und Astronomie, Universität Potsdam, D-14476 Potsdam, Germany}
\email[show]{chuanpeng.hou@uni-potsdam.de}  

\author[0000-0003-2560-8066]{Huirong Yan}
\affiliation{Deutsches Elektronen Synchrotron (DESY), Platanenallee 6, D-15738 Zeuthen, Germany}
\affiliation{Institut für Physik und Astronomie, Universität Potsdam, D-14476 Potsdam, Germany}
\email[show]{huirong.yan@desy.de} 

\author[0000-0003-4268-7763]{Siqi Zhao}
\affiliation{Institut für Physik und Astronomie, Universität Potsdam, D-14476 Potsdam, Germany}
\affiliation{Institute of Science and Technology for Deep Space Exploration, Nanjing University, China}
\email{}

\correspondingauthor{Chuanpeng Hou, Huirong Yan}

\begin{abstract}
Measurements of interplanetary magnetic fields rely mainly on spacecraft observations that provide in-situ sampling and cannot directly reveal global magnetic structures. In addition, the Zeeman effect is generally too weak to image weak interplanetary magnetic fields. Here, we investigate the suitability of spectral-line polarization for spatially resolved imaging of weak magnetic fields, focusing on the Na $\textsc{i}$ D lines in Mercury's magnetosphere. We calculate the spectral-line polarization produced by anisotropic solar radiation and modified by the magnetic-field orientation. Using a three-dimensional forward-imaging model of Mercury's magnetosphere, we generate synthetic maps of Na $\textsc{i}$ D-line intensity and linear polarization. The resulting polarization maps trace the large-scale magnetospheric magnetic-field geometry, demonstrating the potential of Na $\textsc{i}$ D-line polarimetric imaging to constrain the spatial structure of Mercury's magnetosphere. We further show that, with a 1.8-m telescope, Na~$\textsc{i}$ D-line linear-polarization imaging of Mercury's magnetosphere at a spatial resolution of \(2''\) is feasible with an exposure time of approximately ten minutes.
\end{abstract}

\section{Introduction}

Magnetic fields play an important role throughout the heliosphere. They are closely related to solar activity \citep{yang2023global,hou2025fine,liu2025sun}, solar-wind evolution \citep{d2021alfvenic,hou2026non}, and particle transport \citep{yan2022,zhao2025observations,Yan2026}. The interaction between the solar wind and planetary magnetic fields and atmospheres also controls the structure and evolution of planetary magnetospheres and their surrounding plasma environments \citep{mccomas2017jovian,zhao2022observational,chen2023anomalous,exner2024determining,watson2024solar}. Magnetic-field measurements are therefore essential for studies of space and planetary environments \citep{shangguan2013study,zhao2025statistics1,zhao2025statistics2}.

A large number of spacecraft, including Parker Solar Probe, Solar Orbiter, Wind, Cluster, the Magnetospheric Multiscale (MMS) mission, New Horizons, and Voyagers, have measured magnetic fields over a wide range of heliocentric distances \citep{he2015evidence,li2020evolution,rouillard2020relating,dakeyo2022statistical,d2025alfvenic,fraternale2022ssr,fraternale2026,Zhao2021MD,Zhao2022MD,zhao2026mode}. These measurements provide detailed in-situ information, but they are limited to the spacecraft trajectories. Global magnetic-field structures, such as planetary magnetospheres, usually have to be reconstructed by combining measurements from multiple spacecraft and time intervals \citep{tsyganenko2007magnetospheric,wang2021modular}. This limits the ability to obtain magnetic-field images with high temporal resolution.

Several remote-sensing techniques can provide magnetic-field information. Faraday rotation measures the rotation of the polarization plane of radio signals propagating through magnetized plasma \citep{kooi2021vla,kooi2022modern,morgan2023detection}. It can probe regions that are difficult to sample directly with spacecraft \citep{imamura2014outflow,patzold2016mars,wexler2020coronal,ma2022detecting}. However, this technique requires suitable radio sources or spacecraft signals behind the target region. The measurements are also limited to a small number of lines of sight. For specific targets such as Mercury's magnetosphere, radio waves/signals cannot pass through the target region. These limitations make high-resolution two-dimensional imaging difficult. Spectropolarimetric techniques provide another way to diagnose magnetic fields. The Zeeman effect is widely used to measure photospheric magnetic fields \citep{berdyugina2002molecular,chen2025direct}, while the Hanle effect often used in weaker-field regions of the solar atmosphere \citep{del2016magnetic,li2024mapping}. Coronal magnetic fields can also be inferred through coronal seismology \citep{yang2020global}, among other techniques. However, magnetic fields in interplanetary space and planetary environments can be even weaker than those in the solar atmosphere, such that the Hanle effect becomes insensitive to the field strength. Imaging such weak magnetic fields therefore remains challenging.

Spectral-line polarization from anisotropic radiation offers a possible diagnostic for these weak-field environments. The observed polarization can receive contributions from both the lower- and upper-level atomic polarization. In the weak-field interplanetary environments considered here, the polarization is mainly modified by the magnetic-field orientation through ground-state alignment (GSA) \citep{landi2004polarization,yan2006polarization,yan2007polarization,yan2008atomic,yan2012,shangguan2013study,zhang2018influence,zhang2020discovery}. In particular, \cite{shangguan2013study} showed that sub-Gauss interplanetary magnetic fields may be traced using the polarization of the Na~$\textsc{i}$ D$_2$ line.

In this work, we explore the viability of spectral-line polarization for imaging magnetic fields in Mercury's magnetosphere. We forward model the Na~$\textsc{i}$ D-line intensity and linear polarization, and the results show that spectral-line polarization can provide global information on the magnetospheric magnetic field. This paper is organized as follows.
Section \ref{sec2} introduces the basic physics of spectral-line polarization. Section \ref{sec3} presents the forward modeling of the Na~$\textsc{i}$ D lines in Mercury's magnetosphere and the main results. Section \ref{sec4} discusses the observational feasibility of Na~$\textsc{i}$ polarization measurements, and Section \ref{sec5} summarizes the results of this work.

\begin{figure*}[ht]
\centering
\includegraphics[width=0.5\textwidth]{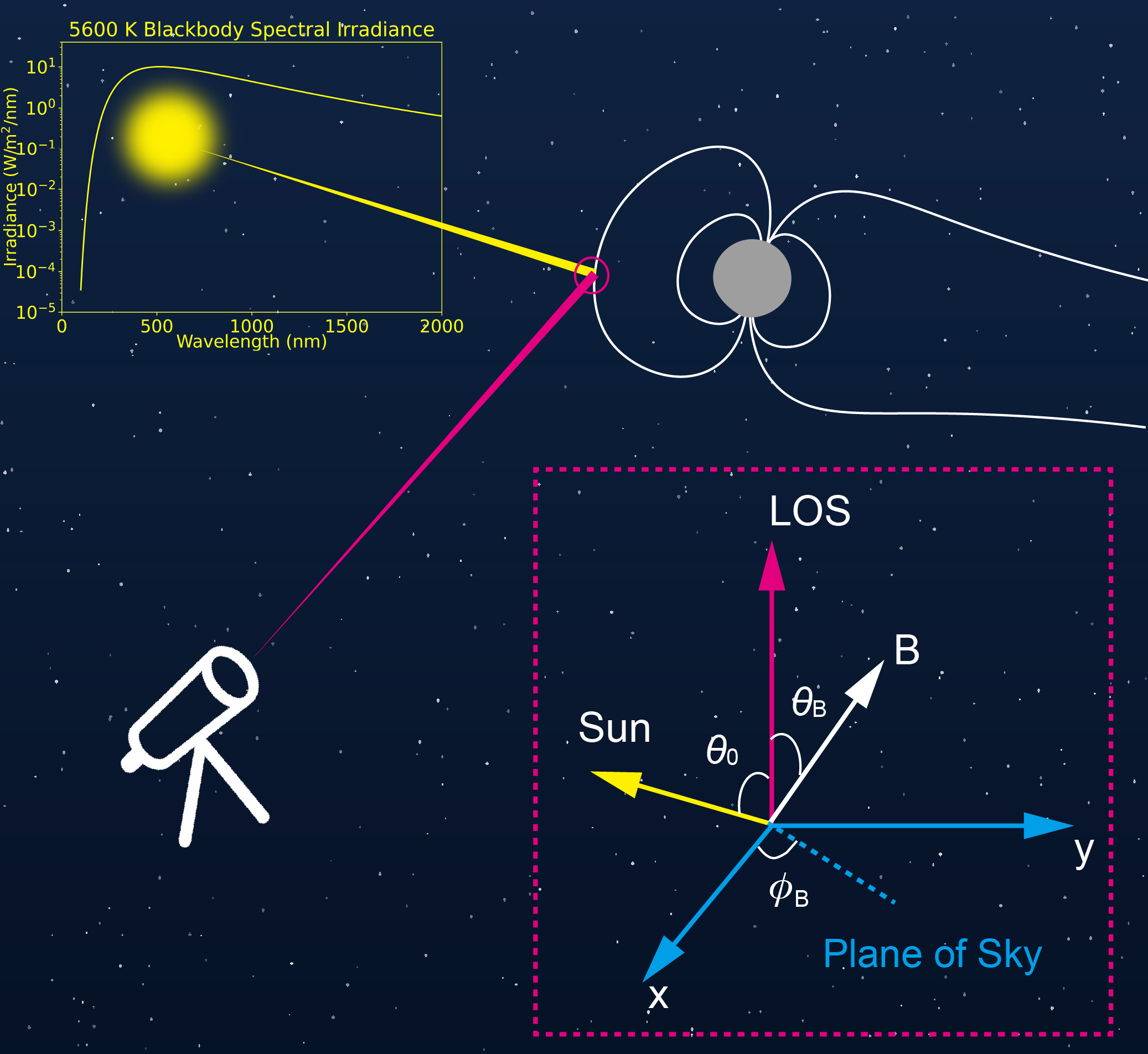}
\caption{Schematic overview of the geometry for the interplanetary magnetic field imaging method. Solar radiation serves as the pumping source, while emission and absorption spectral lines carry polarization signatures that depend on the magnetic field. The angle $\theta_0$ denotes the angle between the incident radiation and the LOS. The magnetic field direction is parameterized by the spherical angles $\theta_B$ and $\phi_B$. }\label{fig1} 
\end{figure*}

\section{Spectral line polarization for magnetic-field diagnostics}\label{sec2}

\subsection{Ground-state alignment and magnetic-field orientation}\label{sec2.1}

The magnetic field imaging method is based on the fact that, when the ground or metastable energy state of an atom contains multiple magnetic sublevels, anisotropic radiation pumping produces unequal sublevel populations, leading to ground state alignment (GSA) \citep{yan2006polarization,yan2007polarization,yan2014magnetic}. This alignment changes the absorption and emission properties of the transition between energy levels, leading to polarized spectral lines. In the presence of a magnetic field, when measuring the alignment with the quantization axis for total angular momentum taken parallel to the magnetic field, the coherence is zero if the Larmor frequency substantially exceeds the radiative pumping rate. In this case, the system is in Hanle saturation, and therefore the polarization depends on the radiation geometry and the magnetic field direction. Therefore, the magnetic field vectors can be inferred from spectral-line linear polarization, enabling magnetic-field imaging of the target (Figure \ref{fig1}).

Quantitatively, the method requires solving the steady-state population equations, including radiative pumping, spontaneous emission, and collisions (Appendices \ref{appendix_fine} and \ref{appendix_hyper}). 
At steady state, the balance of transitions among energy levels allows us to formulate a density-matrix for the population distributions. By solving the density-matrix Equations (\ref{fine_ul}) and (\ref{hyper_ul}), we obtain the population distributions $\rho_q^k(J)$ and $\rho_q^k(F)$ over all relevant energy levels. Here, $\rho_q^k(J)$ and $\rho_q^k(F)$ are irreducible components of the atomic density matrix, describing level populations, population imbalances, and coherences. J is the total electronic angular momentum, while F is the total angular momentum including nuclear spin. The indices k and q denote the tensor rank and component, respectively. For the spectral-line transition between a given pair of energy levels, these quantities are used to calculate the Stokes-parameter absorption ($\eta_I$, $\eta_Q$, $\eta_U$ from Equations (\ref{absorption_fine}) and (\ref{absorption_hyper})) and emission coefficients ($\epsilon_I$, $\epsilon_Q$, $\epsilon_U$ from Equations (\ref{appendix_emission_fine}) and (\ref{appendix_emission_hyper})). The Stokes Q and U are defined with respect to a reference axis aligned with the magnetic field.
The linear polarization degree and angle are generally defined as $p=\sqrt{Q^2+U^2}/I$ and $\chi=\frac{1}{2}\operatorname{arctan}(U/Q)$, respectively.

\begin{figure*}[ht]
\centering
\includegraphics[width=1.0\textwidth]{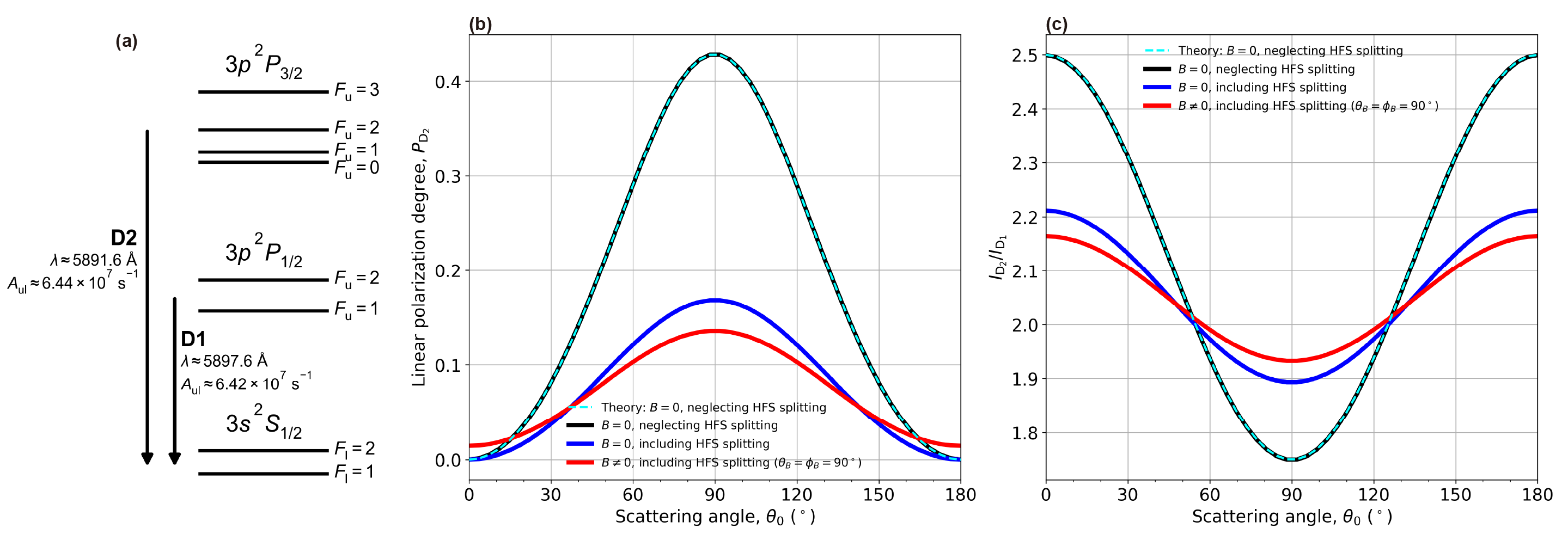}
\caption{Energy-level diagram of the Na~\textsc{i} D lines and the effects of the magnetic field on D-line polarization. (a) Energy-level structure of the Na~\textsc{i} D lines, where $A_{ul}$ is the Einstein coefficient from the upper level to the lower level. (b) Na~\textsc{i} $D_2$ linear polarization as a function of the scattering angle ($\theta_{0}$). The cyan dashed curve shows the theoretical polarization for $B=0$ with HFS neglected. The black solid curve shows the corresponding result by solving density-matrix. The blue solid curve shows the density-matrix result for $B=0$ with HFS included, while the red solid curve shows the result with both HFS and the magnetic field included. (c) Same as panel (b), but for the intensity ratio $I_{\rm D_2}/I_{\rm D_1}$.
}\label{fig2} 
\end{figure*}

Figure \ref{fig2}(a) shows the energy-level structure of the Na~$\textsc{i}$ D lines. The lower level is $3s\,^2S_{1/2}$, while the upper levels are $3p\,^2P_{1/2}$ and $3p\,^2P_{3/2}$. When the nuclear spin is included, these levels are split into eight hyperfine levels. Figures~\ref{fig2}(b) and \ref{fig2}(c) compare the linear polarization and the $I_{D_2}/I_{D_1}$ intensity ratio for different cases, showing the effects of the magnetic field and hyperfine structure (HFS). For all cases, the incident solar radiation is assumed to be unpolarized. In the case where HFS is neglected and $B=0$, both the $D_2$ linear polarization and the intensity ratio depend only on the scattering angle, $\theta_0$, and can be calculated as follows (derived from Equations~(2) and (4) of \citet{langowski2016retrieval} and Equation~(24) of \citet{trujillo2017physics}): 
\begin{equation}
P_{\rm D_2}
=
\frac{3\sin^2\theta_0}
{7+3\cos^2\theta_0},\qquad
\frac{I_{\rm D_2}}{I_{\rm D_1}}
=
\frac{7+3\cos^2\theta_0}{4}.
\end{equation}

The results obtained from our density-matrix calculation agree with these analytical expressions (Figures \ref{fig2}(b) and \ref{fig2}(c)). When HFS is included and $B=0$, the Na~$\textsc{i}$ $D_2$ line is depolarized. A finite magnetic field ($B\ne0$) further modifies both the polarization degree and the intensity ratio. In all cases, the polarization degree and the intensity ratio are anti-correlated.

\subsection{Magnetic-field strength sensitivity in the Hanle regimes}\label{sec2.2}

\begin{figure*}[hb]
\centering
\includegraphics[width=1.0\textwidth]{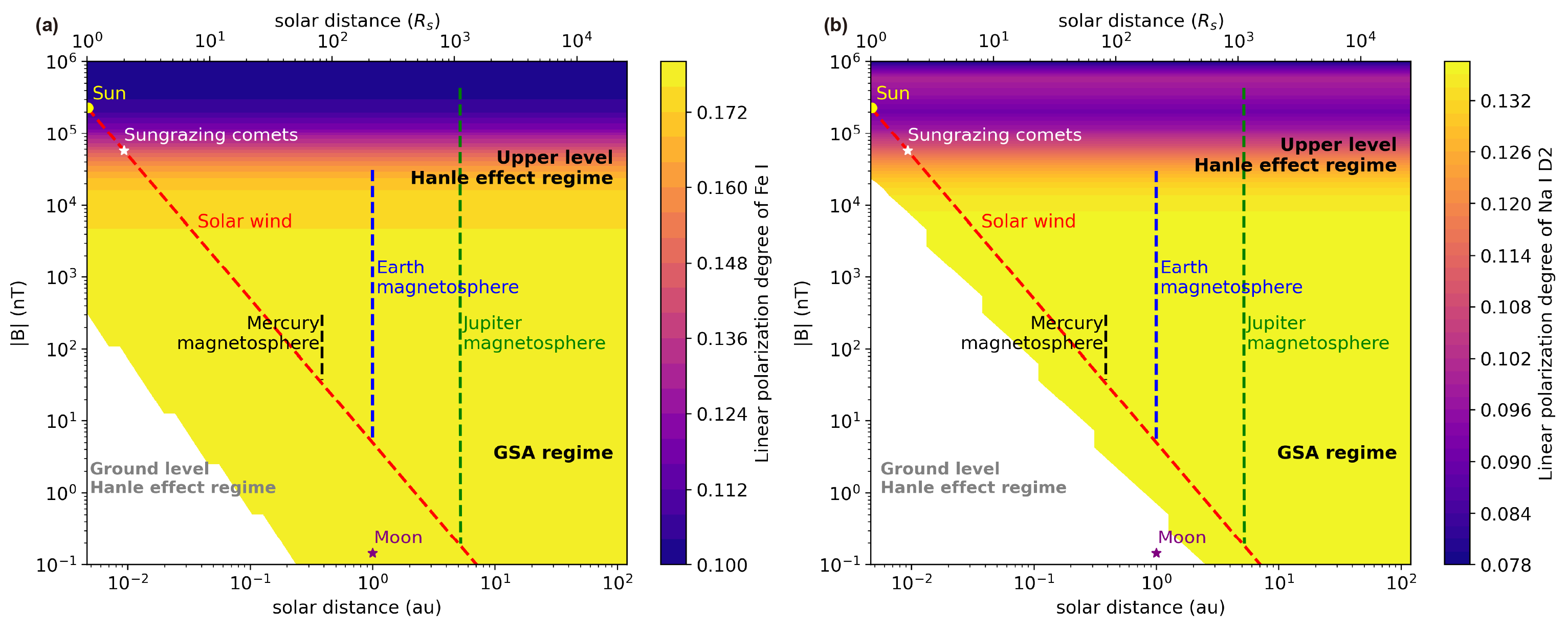}
\caption{The polarization degree as a function of magnetic field strength and solar distance. The observation geometry is $\theta_0=\theta_B=\phi_B=90^\circ$. (a) Fe~\textsc{i} emission line with a wavelength of 3719~\AA. The blank region in the lower-left corner means the ground-level Hanle effect regime ($2\pi\nu_L \le B_{lu}J_0$). (b) Na~\textsc{i} $D_2$ line emission. The dashed lines represent the typical radial profile of magnetic field strength across different environments. The effective blackbody temperature of the Sun is assumed to be 5600~$\mathrm{K}$.
}\label{fig3} 
\end{figure*}

By comparing the relevant rates, we can divide the system into three regimes. When the Larmor precession rate ($\nu_L$) of the electron is comparable to the radiative pumping rate of ground levels ($2\pi\nu_L \sim B_{lu}\bar J^0_0$), the system is in the ground-level Hanle regime.
When the Larmor precession rate is much larger than the radiative pumping rate but still much smaller than the Einstein A coefficient of upper levels ($B_{lu}\bar J^0_0\ll 2\pi\nu_L \ll A$), the system is in the GSA regime, as introduced in section \ref{sec2.1}. When the background magnetic field is sufficiently strong that the Larmor precession rate becomes comparable to the Einstein A coefficient of the upper levels ($2\pi\nu_L \sim A$), the system is in the upper-level Hanle effect regime and the degree of polarization becomes sensitive to the magnetic field strength. In Figure \ref{fig3}, we show which magnetic field strength ranges correspond to each of the three regimes, for cases of the Fe $\textsc{i}$ 3719~\AA\ and the Na $\textsc{i}$ $D_2$ spectral lines. It shows that the linear polarization can be divided into three regimes: the ground-level Hanle regime, selected by $2\pi\nu_L \le B_{lu}J_0$ and marked as blank; the GSA regime, in which the linear polarization degree is independent of $|B|$; and the upper-level Hanle regime, in which the linear polarization degree varies with $|B|$.

For the two lines, the figure shows these regimes for magnetic field strengths typical of different heliospheric environments. The magnetic field in solar wind decreases approximately as $r^{-2}$ with heliocentric distance and is about 5~nT at 1~au. In addition, planets with intrinsic magnetic dipole moments generate localized regions with enhanced magnetic field. For example, the equatorial surface magnetic field is about 300~nT for Mercury, $3.1\times 10^4$~nT for Earth, and $4.28\times 10^5$~nT for Jupiter. In the planetary magnetospheres, these dipolar fields are stronger than that of the surrounding solar wind, but they decrease with distance as approximately $r^{-3}$. By contrast, comets generally lack an intrinsic magnetic field, so the magnetic field in cometary environments can be approximated by the background solar wind. 

Figures~\ref{fig3}(a) and \ref{fig3}(b) show that, in most interplanetary environments, the magnetic field is too weak for the upper-level Hanle effect to be important for the Fe~\textsc{i} 3719~\AA\ and Na~\textsc{i} $D_2$ lines. Their polarization therefore remains mainly in the GSA regime. The upper-level Hanle effect becomes relevant only in stronger-field regions, such as Jupiter's magnetosphere and the high solar atmosphere within about $10 R_{\odot}$. The relatively high Hanle critical magnetic fields of these two lines are related to their large Einstein $A$ coefficients, $1.62\times10^{7}\ {\rm s}^{-1}$ for Fe~\textsc{i} 3719~\AA\ and $6.44\times10^{7}\ {\rm s}^{-1}$ for Na~\textsc{i} $D_2$.
For comparison, the S~\textsc{iii} 1729~\AA\ line has a much smaller Einstein $A$ coefficient of $1.54\times10^{4}\ {\rm s}^{-1}$ and enters the Hanle regime at magnetic fields of order $100$~nT. However, its small $A$ coefficient also leads to weak emission, making high-quality polarization measurements more difficult. Therefore, in interplanetary environments, it is generally easier to determine the magnetic-field orientation than its strength.

\subsection{Collisional effects on atomic alignment}\label{sec2.3}

Collisions tend to reduce atomic alignment by redistributing populations among magnetic sublevels and damping the coherences between them. In the present treatment, we include collisional transitions among ground sublevels and collisional excitation from the ground sublevels to the upper levels. The treatment of collisions involving the upper levels depends on whether hyperfine structure is included. For atoms treated with fine structure only, the radiative lifetimes of the upper levels are much shorter than the relevant collisional timescales. We therefore neglect collisional transitions among upper sublevels and collisional de-excitation from the upper to lower levels. The corresponding collisional terms are given in Appendix \ref{appendix_fine}.

For atoms with nonzero nuclear spin $I$, the electronic angular momentum $J$ couples to $I$, leading to the total angular momentum $F$, and each fine-structure level is split into hyperfine levels (see Figure \ref{fig2}(a)). Because the hyperfine energy separations are small, coherences between different hyperfine levels must also be taken into account, introducing off-diagonal density-matrix elements such as $\rho_q^k(F_u,F_u')$. In this hyperfine-structure treatment, we additionally include collisional transitions among the upper hyperfine levels and collisional deexcitation from the upper to lower levels. The full equations are presented in Appendix \ref{appendix_hyper}.

\subsection{Potential spectral-line polarization in different heliospheric environments}\label{sec2.4}

Table~\ref{table1} summarizes several atomic and ionic spectral lines that may be used for polarization diagnostics in different heliospheric environments. For each species, we select ground levels that are expected to be significantly populated under realistic environments and include all upper levels connected to them through allowed electric-dipole (E1) transitions. Omitting relevant upper levels, particularly those associated with transitions having large Einstein A coefficients, can affect the calculated polarization degree. The wavelength ranges listed in Table~\ref{table1} therefore include all transitions between the selected lower levels and their connected upper levels and generally encompass multiple multiplets.
Tables~\ref{table2} and \ref{table3} summarize the polarization properties of emission and absorption lines with strong polarizability. For the convenience of the research community, we provide an online tool\footnote{\url{https://www.path.physik.uni-potsdam.de/gsa_calculator/}} for calculating the spectral-line linear polarization under any magnetic field direction and strength within the GSA and upper-level Hanle regimes.

The availability of these lines depends strongly on the local atomic and ionic composition. For example, the Na~\textsc{i} D lines have been used for polarization diagnostics in the solar atmosphere \citep{alsina2021solving}. Depending on the local magnetic-field strength, their polarization may constrain the field orientation through GSA and, in the upper-level Hanle regime, provide information on the field strength. Sungrazing comets can supply Fe and Ca species to the upper solar atmosphere \citep{thackeray1966daytime,preston1967spectrum,bida2017observations,fulle2007discovery,harker2023dust,sun2023highlight,raymond2018comet}, while highly ionized species such as C~\textsc{iv} are naturally present there \citep{landi2012carbon}. In the Jovian magnetosphere, oxygen and sulfur ions are abundant and occur in several charge states \citep{clark2016charge,sun2023highlight,2024ApJ...972..154C}. These differences in composition provide a broad range of candidate spectral lines for polarization measurements across heliospheric environments. The observation using these spectral lines, however, also depends on observational detectability. A sufficiently high number density is required to produce detectable line emission, and polarimetric measurements at shorter ultraviolet and extreme ultraviolet wavelengths are more challenging because of the lower photon flux and the stricter requirements on optical components, detectors, and polarization modulation \citep{iglesias2019instrumentation}. Therefore, the applicability of these spectral lines depends on the observational instruments and the specific physical conditions of each environment.

\begin{table*}[hb]
\tabletypesize{\scriptsize}
\setlength{\tabcolsep}{2.5pt}

\begin{ruledtabular}
\caption{Spectral lines from different atoms and ions.}
\label{table1}
\begin{tabular}{@{}llccc@{}}
\cellstack{Species} &
\cellstack{Lower levels $^{a}$} &
\cellstack{Wavelength range of $^{a}$\\included transitions (\AA)} &
\cellstack{Upper-level Hanle $^{b}$\\range (nT)} &
\cellstack{Lower-level Hanle range (nT) $^{b,~c}$\\at $T=5600$~$\mathrm{K}$ and $r=86 ~R_{\rm sun}$}
\\
\midrule

\cellstack[l]{Na~\textsc{i}} & \cellstack[l]{$3s\,{}^{2}S_{1/2}$} &  \cellstack{5891--5898} &  \cellstack{$10^4$--$10^6$}& \cellstack{$< 3\times10^{0}$}
\\

\cellstack[l]{O~\textsc{i}} & \cellstack[l]{$2p^4\,{}^{3}P_{0,1,2}$} & \cellstack{770--1307} & \cellstack{$10^5$--$10^7$} & \cellstack{$< 10^{-6}$}
\\

\cellstack[l]{O~\textsc{ii}} & \cellstack[l]{$2p^3\,{}^{4}S^{\circ}_{3/2}$} & \cellstack{832--835} & \cellstack{$10^5$--$10^7$} & \cellstack{$< 10^{-8}$}
\\

\cellstack[l]{O~\textsc{iii}} & \cellstack[l]{$2p^2\,{}^{3}P_{0,1,2}$} & \cellstack{475--1667} & \cellstack{$10^{-1}$--$10^1$} & \cellstack{$< 10^{-7}$}
\\

\cellstack[l]{S~\textsc{i}} & \cellstack[l]{$3p^4\,{}^{3}P_{0,1,2}$} & \cellstack{1195--1915} & \cellstack{$10^1$--$10^7$}& \cellstack{$< 10^{-6}$}
\\

\cellstack[l]{S~\textsc{ii}} & \cellstack[l]{$3p^3\,{}^{4}S^{\circ}_{3/2}$} & \cellstack{1250--6731} &
\cellstack{$10^4$--$10^6$} & \cellstack{$< 10^{-7}$}
\\

\cellstack[l]{S~\textsc{iii}} & \cellstack[l]{$3p^2\,{}^{3}P_{0,1,2}$} & \cellstack{677--1728} & \cellstack{$10^1$--$10^2$} & \cellstack{$< 10^{-7}$}
\\

\cellstack[l]{S~\textsc{iv}} & \cellstack[l]{$3p\,{}^{2}P^{\circ}_{1/2,3/2}$} & \cellstack{551--1424} &  \cellstack{$10^0$--$10^3$} & \cellstack{$< 10^{-7}$}
\\

\cellstack[l]{Fe~\textsc{i}} & \cellstack[l]{$4s^2\,a\,{}^{5}D_{0,1,2,3,4}$} & \cellstack{1934--5255} & \cellstack{$10^0$--$10^6$} & \cellstack{$< 10^{-2}$}
\\

\cellstack[l]{Fe~\textsc{ii}} & \cellstack[l]{$4s\,a\,{}^{6}D_{1/2,3/2,5/2,7/2,9/2}$\\$4s^2\,a\,{}^{6}S_{5/2}$} & \cellstack{926--5170} & \cellstack{$10^5$--$10^6$} & \cellstack{$< 10^{-1}$}
\\

\cellstack[l]{Ca~\textsc{ii}} & \cellstack[l]{$4s\,{}^{2}S_{1/2}$\\$3d\,{}^{2}D_{3/2,5/2}$} & \cellstack{3934--3969\\8498--8662} &  \cellstack{$10^5$--$10^6$} & \cellstack{$< 10^{0}$}
\\

\end{tabular}

\par\noindent\\[-2pt]
$^{a}$ Atomic-level and transition are obtained from the
\href{https://www.nist.gov/pml/atomic-spectra-database}
{NIST Atomic Spectra Database}. Some transitions are also listed in \citet{yan2006polarization}.\\
$^{b}$ The listed magnetic-field strengths are approximate. More precise values for individual spectral lines are available from our web tool.\\
$^{c}$ $T$ is the solar blackbody radiation temperature, and $r$ is the heliocentric distance with a unit of solar radius ($R_{\rm sun}$).

\end{ruledtabular}
\end{table*}

\begin{table*}[hb]
\tabletypesize{\scriptsize}
\setlength{\tabcolsep}{2.5pt}

\begin{ruledtabular}
\caption{Polarization properties of emission lines from different atoms and ions.}
\label{table2}

\begin{tabular}{@{}lcccccc@{}}

\cellstack{Species} &
\cellstack{Lower level$^{a}$} &
\cellstack{Upper level$^{a}$} &
\cellstack{$\lambda$ (\AA)$^{a}$} &
\cellstack{$A_{ul}$ ($\mathrm{s}^{-1}$)$^{a}$} &
\cellstack{Polarizability$^{b}$} &
\cellstack{$P_{\rm max}$$^{c}$}
\\

\midrule

\cellstack[l]{Na~\textsc{i}} & \cellstack[l]{$3s\,{}^{2}S_{1/2}$} &  \cellstack[l]{$3p\,{}^{2}P^{\circ}_{3/2}$} & \cellstack{5891.6} & \cellstack{$6.44\times 10^7$} & \cellstack{0.5} & \cellstack{0.168}
\\

\cellstack[l]{O~\textsc{i}} & \cellstack[l]{$2p^4\,{}^{3}P_{0}$} & \cellstack[l]{$3s\,{}^{3}S^{\circ}_{1}$} & \cellstack{1306.0} & \cellstack{$6.76\times10^7$} & \cellstack{1.0} & \cellstack{0.332}
\\

\cellstack[l]{O~\textsc{ii}} & \cellstack[l]{$2p^3\,{}^{4}S^{\circ}_{3/2}$} & \cellstack[l]{$2p^4\,{}^{4}P_{3/2}$} & \cellstack{833.3} & \cellstack{$8.65\times10^8$} & \cellstack{0.32} & \cellstack{0.306}
\\

\cellstack[l]{O~\textsc{iii}} & \cellstack[l]{$2p^2\,{}^{3}P_{0}$} & \cellstack[l]{$2p^3\,{}^{3}D^{\circ}_{1}$}& \cellstack{832.9} & \cellstack{$3.41\times10^8$} & \cellstack{1.0} & \cellstack{0.153}
\\

\cellstack[l]{S~\textsc{i}} & \cellstack[l]{$3p^4\,{}^{3}P_{0}$} & \cellstack[l]{$4s\,{}^{3}S^{\circ}_{1}$} & \cellstack{1826.2} & \cellstack{$5.64\times10^7$} & \cellstack{1.0} & \cellstack{0.303}
\\

\cellstack[l]{S~\textsc{ii}} & \cellstack[l]{$3p^3\,{}^{4}S^{\circ}_{3/2}$} & \cellstack[l]{$3p^3\,{}^{2}D^{\circ}_{3/2}$} &\cellstack{6730.8} & \cellstack{$6.84\times10^{-4}$} & \cellstack{0.32} & \cellstack{0.311}
\\

\cellstack[l]{S~\textsc{iii}} & \cellstack[l]{$3p^2\,{}^{3}P_{0}$} & \cellstack[l]{$3p^3\,{}^{3}D^{\circ}_{1}$} & \cellstack{1190.2} & \cellstack{$4.05\times10^7$} & \cellstack{1.0} & \cellstack{0.320}
\\

\cellstack[l]{S~\textsc{iv}} & \cellstack[l]{$3p\,{}^{2}P^{\circ}_{1/2}$} & \cellstack[l]{$3p^2\,{}^{4}P_{3/2}$} & \cellstack{1398.0} & \cellstack{$1.22\times10^3$} & \cellstack{0.5} & \cellstack{0.429}
\\

\cellstack[l]{Fe~\textsc{i}} & \cellstack[l]{$4s^2\,a\,{}^{5}D_{0}$} & \cellstack[l]{$4p({}^{3}P)\,z\,{}^{7}D_{1}$} & \cellstack{5250.2} & \cellstack{$9.30\times10^2$} & \cellstack{1.0} & \cellstack{0.104}
\\

\cellstack[l]{Fe~\textsc{ii}} & \cellstack[l]{$4s\,a\,{}^{6}D_{1/2}$} & \cellstack[l]{$4p\,z\,{}^{6}D^{\circ}_{3/2}$} & \cellstack{2628.3} & \cellstack{$8.74\times10^7$} & \cellstack{0.5} & \cellstack{0.213}
\\

\cellstack[l]{Ca~\textsc{ii}} & \cellstack[l]{$4s\,{}^{2}S_{1/2}$} & \cellstack[l]{$4p\,{}^{2}P^{\circ}_{3/2}$} & \cellstack{3933.7} & \cellstack{$1.47\times10^8$} & \cellstack{0.5} & \cellstack{0.396}
\\

\end{tabular}

\par\noindent\\[-2pt]
$^{a}$ Energy levels and transitions are obtained from the
\href{https://www.nist.gov/pml/atomic-spectra-database}
{NIST Atomic Spectra Database}.\\
$^{b}$ Polarizability of the emission line, defined as
$W_{2, \mathrm{em}} \equiv |\omega^{(2)}_{J_uJ_l}|^2$, where the expression for
$\omega^{(2)}_{J_uJ_l}$ is given in Appendix~A. For Na~\textsc{i}, the polarizability refers to the
fine-structure D$_2$ transition without hyperfine structure.\\
$^{c}$ The maximum linear polarization degree, calculated with $\theta_0=\theta_B=90^\circ$, $\phi_B=0^\circ$. At short wavelengths, the blackbody spectrum underestimates the solar irradiance, and we therefore assume $I_*(\lambda<1000~\mathrm{\AA})=I_*(1000~\mathrm{\AA})$.
\end{ruledtabular}
\end{table*}

\begin{table*}[hb]
\tabletypesize{\scriptsize}
\setlength{\tabcolsep}{2.5pt}

\begin{ruledtabular}
\caption{Polarization properties of absorption lines from different atoms and ions.}
\label{table3}

\begin{tabular}{@{}lcccccc@{}}

\cellstack{Species} &
\cellstack{Lower level$^{a}$} &
\cellstack{Upper level$^{a}$} &
\cellstack{$\lambda$ (\AA)$^{a}$} &
\cellstack{$A_{ul}$ ($\mathrm{s}^{-1}$)$^{a}$} &
\cellstack{Polarizability$^{b}$} &
\cellstack{$P_{\rm max}$$^{c}$}
\\

\midrule

\cellstack[l]{Na~\textsc{i}} & \cellstack[l]{$3s\,{}^{2}S_{1/2}$} &  \cellstack[l]{$3p\,{}^{2}P^{\circ}_{3/2}$} & \cellstack{5891.6} & \cellstack{$6.44\times 10^7$} & \cellstack{0.0} & \cellstack{0.0}
\\

\cellstack[l]{O~\textsc{i}} & \cellstack[l]{$2p^4\,{}^{3}P_{2}$} & \cellstack[l]{$3s\,{}^{3}S^{\circ}_{1}$} & \cellstack{1302.2} & \cellstack{$3.41\times10^8$} & \cellstack{0.35} & \cellstack{0.304}
\\

\cellstack[l]{O~\textsc{ii}} & \cellstack[l]{$2p^3\,{}^{4}S^{\circ}_{3/2}$} & \cellstack[l]{$2p^4\,{}^{4}P_{1/2}$} & \cellstack{832.8} & \cellstack{$8.67\times10^8$} & \cellstack{0.5} & \cellstack{0.127}
\\

\cellstack[l]{O~\textsc{iii}} & \cellstack[l]{$2p^2\,{}^{3}P_{1}$} & \cellstack[l]{$2p^3\,{}^{3}P^{\circ}_{0}$} & \cellstack{702.8} & \cellstack{$1.83\times10^9$} & \cellstack{1.0} & \cellstack{0.248}
\\

\cellstack[l]{S~\textsc{i}} & \cellstack[l]{$3p^4\,{}^{3}P_{1}$} & \cellstack[l]{$3p^5\,{}^{3}P^{\circ}_{0}$} & \cellstack{1385.5} & \cellstack{$1.25\times10^6$} & \cellstack{1.0} & \cellstack{0.444}
\\

\cellstack[l]{S~\textsc{ii}} & \cellstack[l]{$3p^3\,{}^{4}S^{\circ}_{3/2}$} & \cellstack[l]{$3p^3\,{}^{2}P^{\circ}_{1/2}$} &\cellstack{4076.3} & \cellstack{$7.72\times10^{-2}$} & \cellstack{0.5} & \cellstack{0.141}
\\

\cellstack[l]{S~\textsc{iii}} & \cellstack[l]{$3p^2\,{}^{3}P_{1}$} & \cellstack[l]{$3p^3\,{}^{3}P^{\circ}_{0}$} & \cellstack{1015.5} & \cellstack{$2.88\times10^8$} & \cellstack{1.0} & \cellstack{0.020}
\\

\cellstack[l]{S~\textsc{iv}} & \cellstack[l]{$3p\,{}^{2}P^{\circ}_{3/2}$} & \cellstack[l]{$3p^2\,{}^{4}P_{1/2}$} & \cellstack{1423.8} & \cellstack{$4.45\times10^4$} & \cellstack{0.5} & \cellstack{0.168}
\\

\cellstack[l]{Fe~\textsc{i}} & \cellstack[l]{$4s^2\,a\,{}^{5}D_{1}$} & \cellstack[l]{$4p({}^{3}P)\,z\,{}^{5}D_{0}$} & \cellstack{3895.7} & \cellstack{$9.39\times10^6$} & \cellstack{1.0} & \cellstack{0.012}
\\

\cellstack[l]{Fe~\textsc{ii}} & \cellstack[l]{$4s\,a\,{}^{6}D_{3/2}$} & \cellstack[l]{$4p\,z\,{}^{6}D^{\circ}_{1/2}$} & \cellstack{2613.8} & \cellstack{$2.12\times10^8$} & \cellstack{0.5} & \cellstack{0.044}
\\

\cellstack[l]{Ca~\textsc{ii}} & \cellstack[l]{$3d\,{}^{2}D_{3/2}$} & \cellstack[l]{$4p\,{}^{2}P^{\circ}_{1/2}$} & \cellstack{8662.1} & \cellstack{$1.06\times10^7$} & \cellstack{0.5} & \cellstack{0.234}
\\

\end{tabular}

\par\noindent\\[-2pt]
$^{a}$ Energy levels and transitions are obtained from the
\href{https://www.nist.gov/pml/atomic-spectra-database}
{NIST Atomic Spectra Database}.\\
$^{b}$ Polarizability of the absorption line, defined as
$W_{2, \mathrm{ab}} \equiv |\omega^{(2)}_{J_lJ_u}|^2$, where the expression for
$\omega^{(2)}_{J_lJ_u}$ is given in Appendix~A. \\
$^{c}$ The maximum linear polarization degree per unit optical depth ($P=|\eta_Q/\eta_I|$), calculated with $\theta_0=\theta_B=90^\circ$, $\phi_B=0^\circ$. At short wavelengths, the blackbody spectrum underestimates the solar irradiance, and we therefore assume $I_*(\lambda<1000~\mathrm{\AA})=I_*(1000~\mathrm{\AA})$.
\end{ruledtabular}
\end{table*}

For Mercury, the Na~\textsc{i} D lines provide a more favorable case. Sodium is particularly abundant in Mercury's magnetosphere with a column density of ($\sim 10^{11}~\textrm{cm}^{-2}$), and radiation pressure can produce an extended sodium tail reaching millions of kilometers \citep{fei2021development,leblanc2022comparative}. Na~\textsc{i} D-line emission from Mercury has also been detected in previous observations \citep{baumgardner2008imaging}. In addition, resonant scattering in the Na~\textsc{i} D$_2$ line can produce measurable linear polarization \citep{ariste2012resonance}. These properties make the Na~\textsc{i} D lines suitable candidates for polarization measurements in Mercury's magnetosphere. In addition, Fe atoms are also present in Mercury's exosphere, with a zenith column density of $8.2 \times 10^{8}~\mathrm{cm^{-2}}$ \citep{bida2017observations}, we therefore focus on diagnosing the magnetic-field orientation using the Na~\textsc{i} D lines and Fe~\textsc{i} lines in the forward modeling presented in Section~\ref{sec3}.

\section{Forward Modeling of Polarization in Mercury's magnetosphere}\label{sec3}

\subsection{Forward-modeling Procedure}\label{sec3.1}

We perform three-dimensional (3D) forward modeling of Na~\textsc{i} D-line emission and Fe~\textsc{i} 3719~\AA\ absorption by combining a magnetospheric simulation with the atomic-polarization model. The procedure includes five steps: constructing the lines of sight (LOSs), sampling the atomic density and magnetic field along each LOS, calculating the local Stokes emission and absorption coefficients at each sampling point, integrating the emissivities along each LOS, and projecting the results onto a two-dimensional plane of the sky.

\begin{figure*}[ht]
\centering
\includegraphics[width=0.8\textwidth]{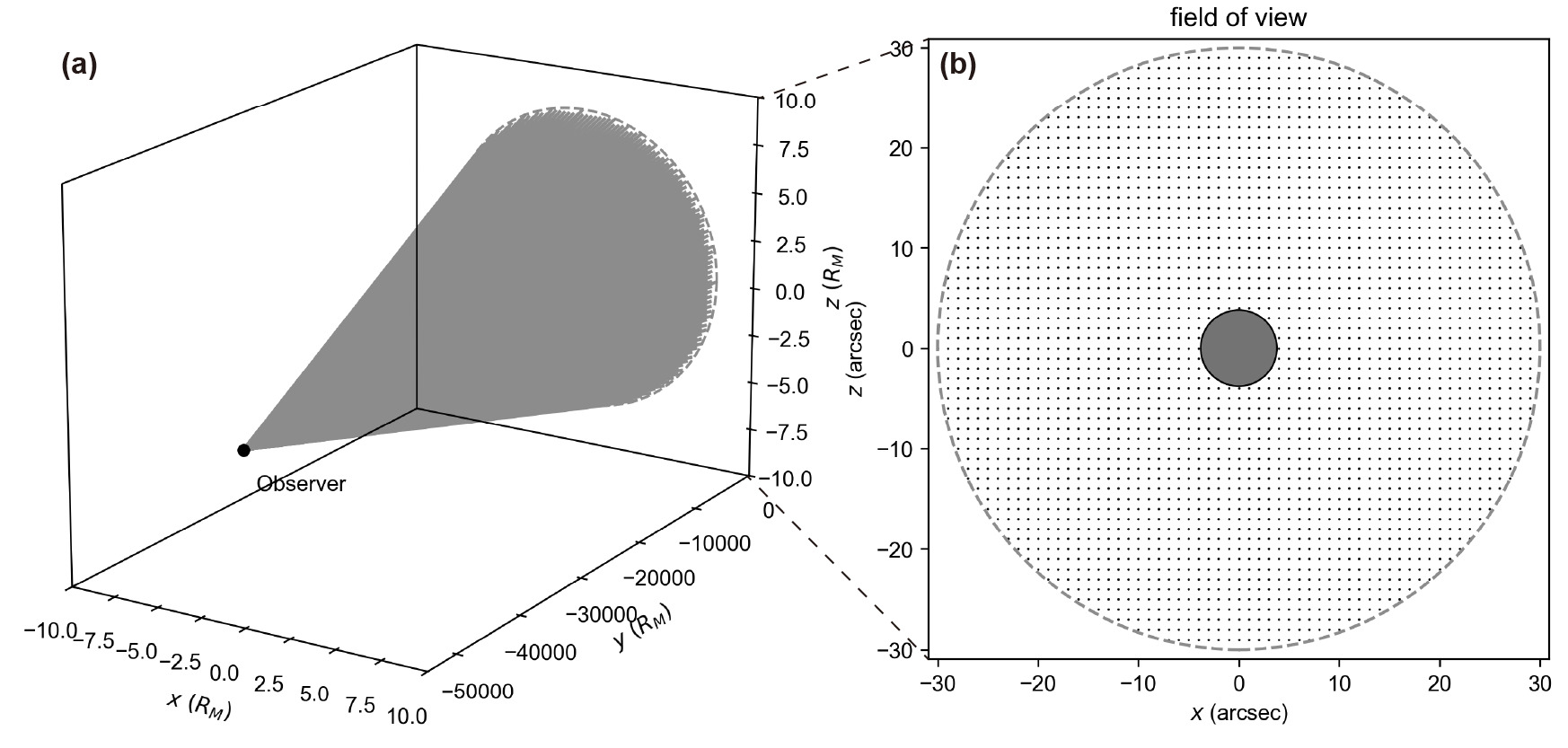}
\caption{Schematic of the field of view used in the forward modeling. (a) Three-dimensional field of view containing all LOS. The LOS extends from the observer toward Mercury and forms a cone. Distances are given in units of Mercury’s radius, $R_{\rm M}$. (b) Two-dimensional projection of the field of view. The angular separation between adjacent LOS determines the angular resolution, and the full set of LOS covers a circular field of view.}\label{fig4} 
\end{figure*}

The observer is placed at Earth, with the central LOS directed toward Mercury. The Sun-Mercury-Earth angle is set to $90^\circ$. We adopt a circular field of view of with a diameter of $60^{\prime\prime}$, with the angular spacing determined by the required image resolution (Figure \ref{fig4}). The LOS integration is limited to $10~R_{\rm M}$ from Mercury, where $R_{\rm M}$ is Mercury's radius. Emission from regions occulted by Mercury is excluded. Each LOS is sampled with 200 uniformly spaced points. The atomic density and magnetic-field components at each sampling point are interpolated from simulation case Sim-1 of the 3D MHD model of Mercury's magnetosphere \citep{chen2024transport,chen_2024_10882126}. At each sampling point, we determine the relative geometry among the incident solar radiation, the local magnetic field direction, and the LOS. These geometrical parameters are then used in the atomic-polarization calculation.

We use a multilevel atomic model for both the fine-structure and HFS calculations. For the lower levels of Fe~\textsc{i} lines, we include all relevant radiatively connected upper levels with allowed electric-dipole (E1) transitions.
Therefore, Fe~\textsc{i} model includes five low-lying lower levels (Table~\ref{table1}), 109 connected upper levels, and 228 transitions in total. These levels and transitions are treated simultaneously in the statistical-equilibrium equations described in Appendix~\ref{appendix_fine}. 
For the Na~\textsc{i} D lines, the atomic model contains three fine-structure levels: the ground level $3s\ ^2S_{1/2}$ and the upper levels $3p\ ^2P_{1/2,3/2}$. When HFS is included, these levels split into eight $F$ levels, producing 10 allowed hyperfine transitions. All these levels and transitions are included in a single density-matrix calculation, described in Appendix~\ref{appendix_hyper}.

After solving the equations for the density matrix, we calculate the local Stokes emissivities $\epsilon_I$, $\epsilon_Q$, and $\epsilon_U$ at each sampling point. Circular polarization is not considered, so $\epsilon_V=0$. The local $Q$ and $U$ components are rotated into a common reference frame before LOS integration. The projection of the simulation $z$-axis onto the image plane is used as the reference direction (Appendix~\ref{appendix_rot}). In these calculations, the incident solar radiation is represented by a blackbody spectrum with an effective temperature of $5600~{\mathrm{K}}$.
We assume that the incident solar radiation is unpolarized. Accounting for the polarization of the incident radiation introduces additional contributions to the radiation-field tensors but does not alter the general framework of the density-matrix and forward-modeling calculations. A complete treatment requires realistic solar Stokes spectra will be considered in future applications to observational data. Collisions are neglected because their rates under Mercury's magnetospheric conditions are much lower than the relevant radiative rates.

The LOS-integrated Stokes parameters for emission lines are calculated as
\begin{equation}
I_{\rm LOS} = \sum_k \epsilon_{I,k}\Delta s_k,\qquad
Q_{\rm LOS} = \sum_k \epsilon_{Q,k}\Delta s_k,\qquad
U_{\rm LOS} = \sum_k \epsilon_{U,k}\Delta s_k,
\end{equation}

and the corresponding linear polarization degree and the polarization angle are
\begin{equation}
P =
\frac{\sqrt{Q_{\rm LOS}^{2}+U_{\rm LOS}^{2}}}
     {I_{\rm LOS}},\qquad
\chi =
\frac{1}{2}
\operatorname{arctan}
\left(U_{\rm LOS}/Q_{\rm LOS}\right).
\label{eq:polarization_angle}
\end{equation}

The LOS-integrated Stokes parameters for absorption lines are calculated as
\begin{equation}
\frac{I_{\rm LOS}}{I_0} = 1-\sum_k \eta_{I,k}\Delta s_k,\qquad
P = \sqrt{(\sum_k \eta_{Q,k}\Delta s_k)^2 + (\sum_k \eta_{U,k}\Delta s_k)^2},\qquad \chi = \frac{1}{2}\arctan(\sum_k \eta_{U,k}\Delta s_k / \sum_k \eta_{Q,k}\Delta s_k),\qquad
\end{equation}

where $\epsilon_{I,Q,U}$ are the emission coefficients for Stokes I, Q, and U, respectively. $\eta_{I}$ is the absorption coefficient for Stokes I, and $\eta_{Q,U}$ are the dichroism coefficients for Stokes Q and U, respectively. $I_0$ is the incident radiation. $\Delta s_k$ is the path length associated with the $k$th sampling point. The atomic column density is calculated as,
\begin{equation}
n = \sum_k n_{k}\Delta s_k.
\end{equation}

The LOS-integrated quantities are finally mapped onto the two-dimensional plane of the sky to produce synthetic maps of intensity, polarization degree, polarization angle, $I_{D_2}/I_{D_1}$ intensity ratio, and atomic column density. For comparison with the polarization angle, we calculate the density-weighted LOS-averaged magnetic field direction as
\begin{equation}
\langle \mathbf{\hat B} \rangle_{\rm LOS}
=
\frac{\sum_k n_{k}\mathbf{B}_k/|\mathbf{B}_k|\Delta s_k}
     {\sum_k n_{k}\Delta s_k},
\end{equation}
where $n_{k}$ and $\mathbf{B}_k$ are the number density and magnetic field at the $k$th sampling point, respectively. The LOS-averaged magnetic field is then projected onto the plane of the sky.

\subsection{Forward-modeling Results}\label{sec3.2}
\subsubsection{Na~\textsc{i} D-line Emission}\label{sec3.2.1}

\begin{figure*}[]
\centering
\includegraphics[width=1.0\textwidth]{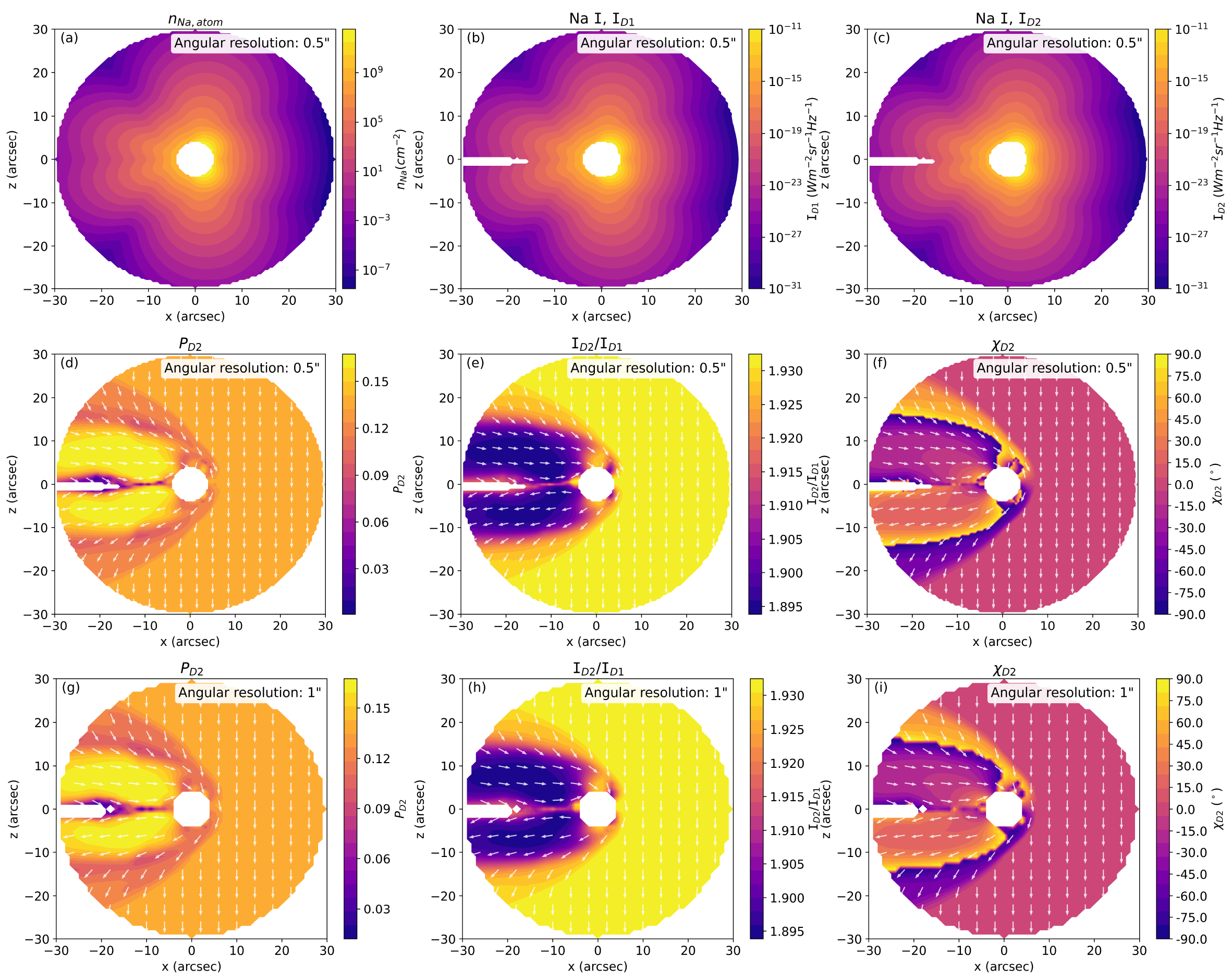}
\caption{Forward-modeled images of the Na emission line linear polarization degree ($P$), intensity ratio ($I_{\rm D_2}/I_{\rm D_1}$), and polarization angle for Mercury’s magnetosphere. (a) LOS-integrated sodium number density. (b–c) Simulated observations of the emission-line intensities for Na $D_1$ and $D_2$ line, respectively. (d-f) Simulated observations at a spatial resolution of $0.5^{\prime\prime}$. (g-i) Simulated observations at a spatial resolution of $1^{\prime\prime}$. The white arrows indicate the density-weighted LOS-averaged magnetic field direction projected onto the plane of the sky. The scattering angle ($\theta_0$), defined as the Sun–Mercury–Earth angle, is approximately $90^\circ$. The polarization measurement along a line of sight is omitted if more than 5\% of sampling points have magnetic-field strengths below 3 nT, corresponding to the lower-level Hanle-effect regime.}\label{fig5} 
\end{figure*}

We present the forward-modeling results for a Sun--Mercury--Earth geometry with a scattering angle of $90^\circ$. The synthetic images have an angular resolution of $0.5^{\prime\prime}$ and a circular field of view with a diameter of $60^{\prime\prime}$. The total number of LOSs is 11104. Collisions reduce the Na~\textsc{i} D-line polarization degree by about 0.07\% at an electron density of approximately $10^{8}~\mathrm{cm}^{-3}$, which is far higher than typical electron densities in Mercury's magnetosphere. Hence, we neglect collisional effects when calculating these images. The temperature of the sodium atoms is assumed to be 1000~$\mathrm{K}$ \citep{leblanc2022comparative,chen2024transport}. The thermal Doppler width of the Na~\textsc{i} D lines is approximately $1.44\times10^9~\mathrm{Hz}$, corresponding to a line-center value of approximately $\xi=3.91\times10^{-10}~\mathrm{Hz^{-1}}$ for the normalized Gaussian profile.

Figure~\ref{fig5}(a) shows the LOS-integrated Na~\textsc{i} density. Figures~\ref{fig5}(b) and \ref{fig5}(c) show the Na~\textsc{i} $D_1$ and $D_2$ intensity maps, respectively. For both lines, the spatial distribution of intensity is the same as Na~\textsc{i} column densities (Figure~\ref{fig5}(a)).
The reason is that at each local position, intensity is proportional to the sodium number density, and the integrated intensity is also proportional to the column density. On the other hand, the intensity distributions do not directly reveal the global magnetospheric geometry. This means that intensity observations of single spectral line is not enough to reveal the magnetic field information.

As shown in Figures~\ref{fig5}(d)--(f), polarization measurements can reveal additional details of Mercury's magnetosphere. At an angular resolution of $0.5^{\prime\prime}$, the Na \textsc{i} D$_2$ polarization degree $P$, the intensity ratio $I_{\rm D_2}/I_{\rm D_1}$, and the polarization angle $\chi$ exhibit spatial patterns that trace the overall morphology of the magnetosphere formed through the interaction between Mercury and the solar wind. The white arrows overlaid on the maps indicate the density-weighted, LOS-averaged magnetic-field direction, $\langle\hat{\boldsymbol{B}}\rangle_{\rm LOS}$, providing a reference for comparing the modeled polarization signals with the magnetic-field geometry. In addition to the global magnetospheric morphology, the polarization maps could also reveal smaller-scale structures associated with the magneto-tail current sheet.

The Na~\textsc{i} D$_2$ polarization degree shown in Figure~\ref{fig5}(d) ranges from approximately $1\%$ to $16\%$. Such relatively strong polarization degree is favorable for polarimetric observations. As discussed in Section \ref{sec4}, under the assumed observational conditions, a polarimetric sensitivity of approximately $1\%$ could be achieved with an exposure time lower to the order of ten minutes. Moreover, the polarization degree in Figure~\ref{fig5}(d) and the intensity ratio $I_{\rm D_2}/I_{\rm D_1}$ in Figure~\ref{fig5}(e) exhibit an overall anti-correlation, consistent with the relation shown in Figure~\ref{fig2}. 

\begin{figure*}[]
\centering
\includegraphics[width=1.0\textwidth]{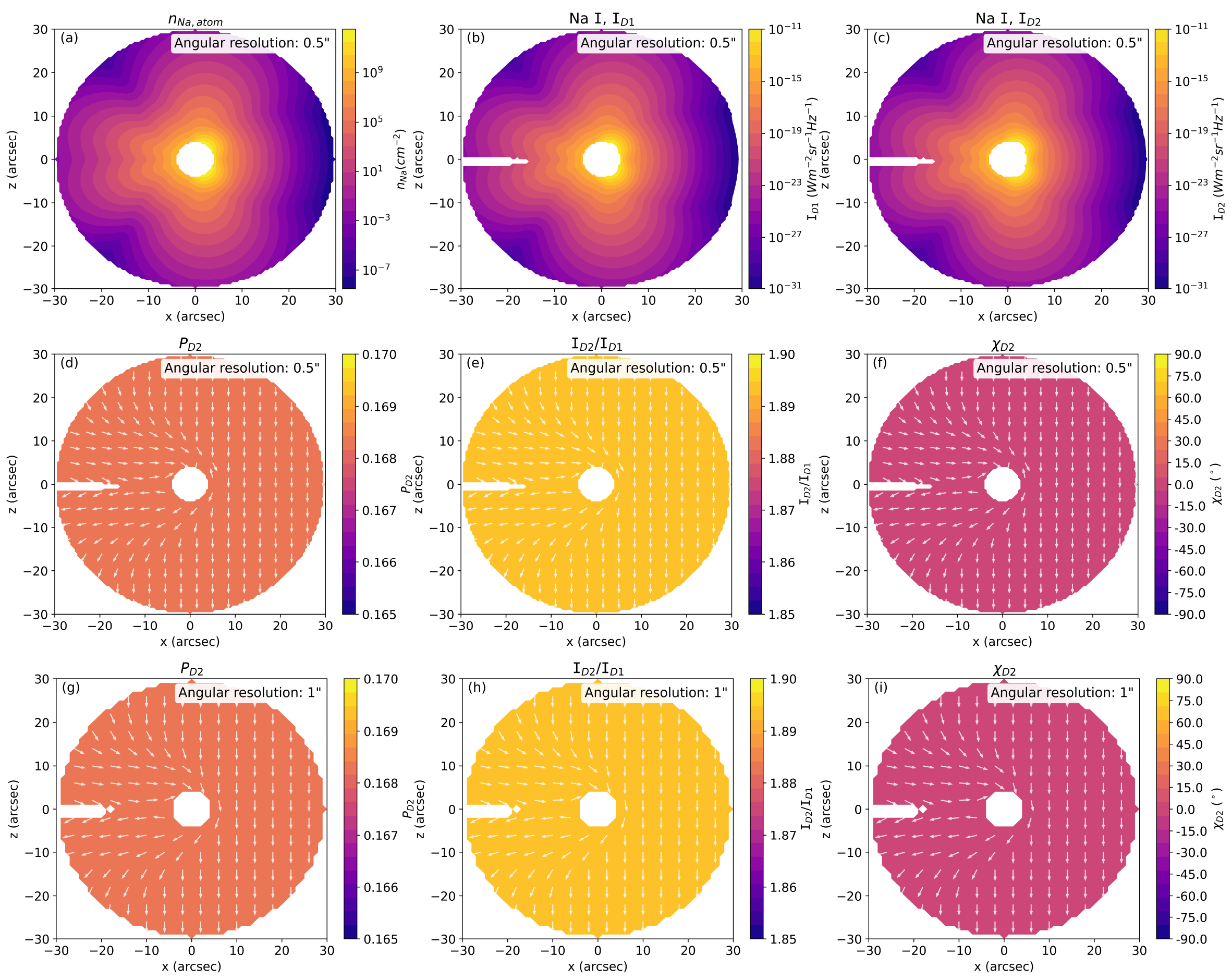}
\caption{Same as Figure~\ref{fig5}, but with no magnetic field ($B=0$).}\label{fig6} 
\end{figure*}

We repeat the forward-modeling calculation at a lower angular resolution of $1^{\prime\prime}$ while keeping the field of view unchanged, resulting in only 2768 LOSs. The same MHD simulation data are used without degrading their original spatial resolution. The forward modeling results are presented in Figures~\ref{fig5}(g)--(i) in the same format as the $0.5^{\prime\prime}$ results. The main magnetospheric structures remain identifiable in Figures~\ref{fig5}(g)--(i). In the absence of a magnetic field (B=0), the polarization degree, polarization angle, and $I_{D_2}/I_{D_1}$ depend only on the scattering geometry. For Mercury's magnetosphere, the scattering angle (Sun-Mercury-Earth angle) is nearly constant. Therefore, these polarization properties will be nearly uniform across the magnetosphere (Figure~\ref{fig6}). These results show that polarization observations with arcsecond-level resolution can provide global information about Mercury's magnetosphere and can be used to study its large-scale structure.

\subsubsection{Fe \textsc{i} Absorption Lines}\label{sec3.2.2}

Compared with emission lines, absorption lines should exhibit a more direct polarization behavior: the absorption for Stokes $U$ is zero with $\eta_{U}=0$, while Stokes $Q$ is either parallel or perpendicular to the magnetic field. Consequently, the polarization angle directly traces the magnetic field orientation projected onto the plane of the sky with a \(90^\circ\) ambiguity. Since, for the Na \textsc{i} D doublet, absorption processes do not introduce linear polarization \citep{yan2007polarization, shangguan2013study}, we need to choose another spectral line. 
For Mercury’s atmosphere, the zenith column density of Fe \textsc{i} is $8.2 \times 10^{8}~\mathrm{cm^{-2}}$ \citep{bida2017observations}. Such density is far lower than the column density of sodium ($\sim 10^{11}~\textrm{cm}^{-2}$) \citep{mcgrath1986sputtering}, and collisions affects are ignored. The Fe \textsc{i} with relatively lower densities are more suitable for imaging the near-Mercury exosphere environment, rather than the extended tail. Absorption-line measurements also require the Sun, Mercury, and the observer to be nearly aligned, as during a Mercury transit. In this geometry, the observer views Mercury approximately along the Sun--Mercury direction from the magneto-tail side.

We perform forward modeling of the Fe~\textsc{i} 3719~\AA\ absorption line. Because a reliable three-dimensional distribution of Fe~\textsc{i} is unavailable, we estimate the Fe~\textsc{i} number density by scaling the Na~\textsc{i} density distribution by a constant factor of $8.2\times10^{-3}$. This constant scaling changes the predicted absorption intensity and polarization degree $P$, but does not affect the polarization angle $\chi$. However, the actual three-dimensional distributions of Fe~\textsc{i} and Na~\textsc{i} may differ, which could modify the spatial polarization patterns obtained from the forward modeling. The adopted scaling factor is therefore used only as a first-order approximation to the Fe~\textsc{i} spatial distribution.

\begin{figure*}[]
\centering
\includegraphics[width=1.0\textwidth]{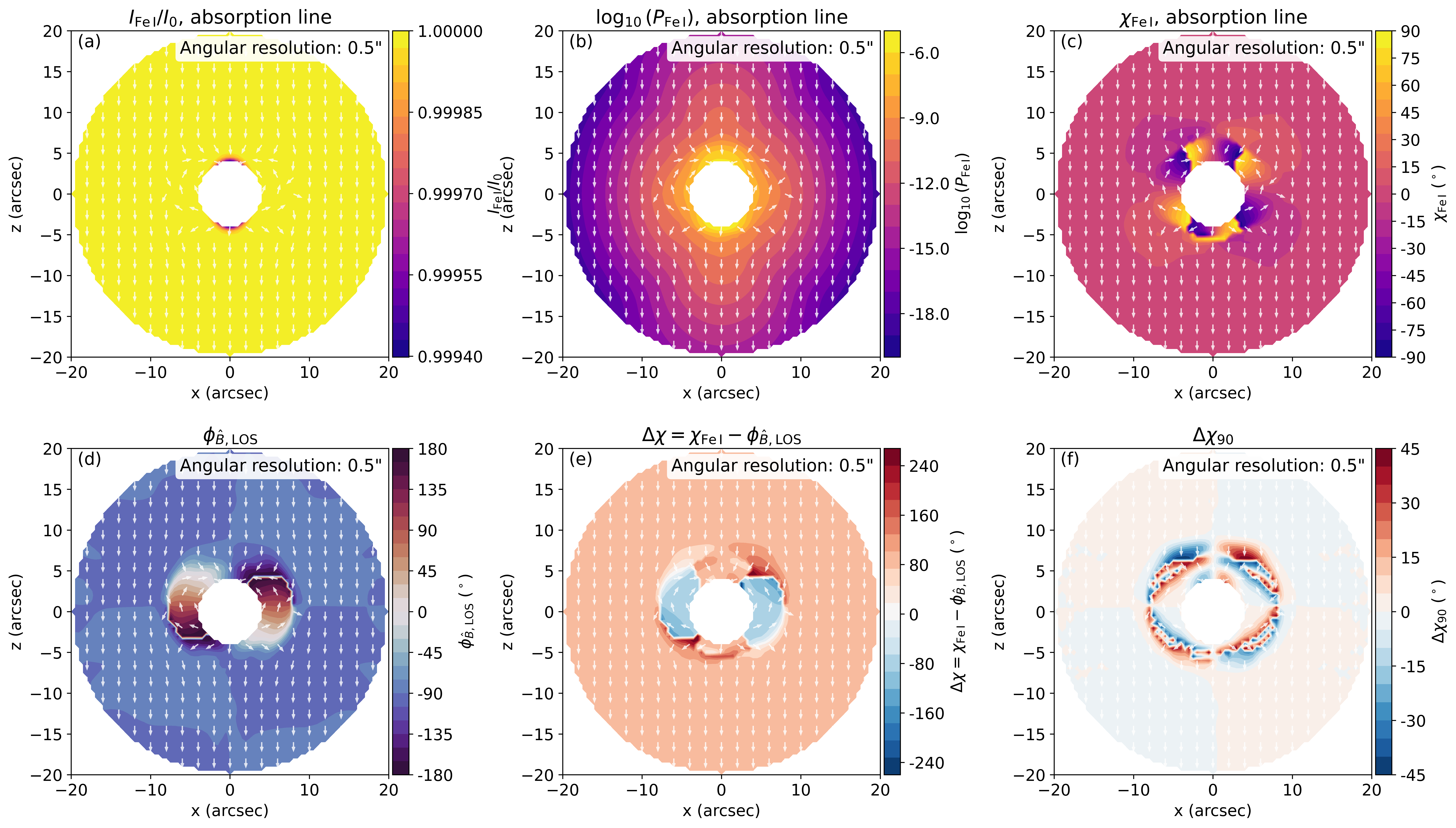}
\caption{Forward-modeled absorption-line images viewed along the LOS from Mercury's magneto-tail toward the Sun. The scattering angle $\theta_0$, defined by the Sun--Mercury--Earth geometry, is approximately $180^\circ$. (a) Line-of-sight map of the Fe~\textsc{i}~3719~\AA\ intensity normalized to the initial intensity \(I_0\). Panels (b) and (c) show the linear polarization degree and polarization angle of the Fe~\textsc{i} 3719~\AA\ line. Panel (d) shows the density-weighted LOS-averaged magnetic-field direction. Panel (e) shows the signed difference between the polarization angle and this magnetic-field orientation. Panel (f) shows the same angular difference folded according to Equation (\ref{chi_90}) to account for the \(90^\circ\) ambiguity.
}\label{fig7} 
\end{figure*}

Figure~\ref{fig7}(a) shows the $I_{\mathrm{Fe}~\textsc{i}}$ intensity normalized by the background radiation intensity ($I_{0}$). The white arrows indicate the density-weighted, LOS-averaged magnetic-field direction projected onto the plane of the sky. Because absorption by the Fe~\textsc{i} line in Mercury's exosphere is weak, the normalized intensity is close to unity over most of the field of view, indicating negligible absorption. The strongest absorption occurs near Mercury's north and south poles, where the observed intensity only decreases to approximately $99.94\%$ of the background intensity. The corresponding linear polarization degree is also weak. Even in the polar regions, the maximum polarization degree is only approximately $10^{-6}$, making its detection impossible in practice. 
The direct angular difference between the polarization angle (Figure~\ref{fig7}(c)) and the projected magnetic-field direction $\phi_{\hat B,LOS}$ (Figure~\ref{fig7}(d)) spans a broad range (Figure~\ref{fig7}(e)), largely because of the inherent \(90^\circ\) ambiguity in the polarization orientation.

To quantify the angular difference while accounting for the \(90^\circ\) ambiguity, we define
\begin{equation}\label{chi_90}
\Delta \chi_{90}
=
\frac{1}{4}\,
\operatorname{atan2}
\left[
\sin\left(4(\chi-\phi_{\rm\hat B,LOS})\right),
\cos\left(4(\chi-\phi_{\rm\hat B,LOS})\right)
\right],
\end{equation}
where $\phi_{\rm\hat B,LOS}$ is the azimuth of the density-weighted LOS-averaged magnetic field direction projected onto the plane of the sky. This definition folds the angular difference into the range \([-45^\circ,~45^\circ]\). A value of $\Delta \chi_{90}=0^\circ$ indicates that the polarization direction is either parallel or perpendicular to the projected magnetic field, whereas \(|\Delta\chi_{90}|=45^\circ\) represents the maximum deviation from the nearest parallel or perpendicular configuration. Figure~\ref{fig7}(f) shows the resulting $\Delta\chi_{90}$. The angular difference is small over most of the field of view but becomes larger close to Mercury, indicating the LOS integration indeed will affect the observation results and make inversions of such magnetic fields difficult. The LOS-integrated polarization depends on the spatial distribution of the absorbing atoms density and magnetic field. Hence, prior atmospheric and magnetic-field models are required to constrain the inversion.

\section{Discussion}\label{sec4}

For spectropolarimetric imaging of Mercury's magnetosphere/exosphere, the Na \textsc{i} emission lines appear to be more promising observational targets than the Fe~\textsc{i}~3719~\AA\ absorption line. Under the modeled conditions, the Na~\textsc{i} D$_2$ line produces a substantially stronger linear-polarization signal, making it more accessible to polarimetric measurements. In contrast, the Fe~\textsc{i} absorption line produces both only a tiny decrease in intensity and a very low polarization degree. Moreover, absorption-line imaging of Mercury can only be performed during a transit across the solar disk. Such events are rare and have limited durations, further restricting the opportunities for absorption-line observations. It would therefore be worthwhile in future work to search for other atomic or ionic absorption lines that exhibit sufficiently strong absorption and polarization for practical observations. At present, Na~\textsc{i} D$_2$ emission-line polarization appears to provide a more practical approach for imaging Mercury's magnetospheric magnetic-field structure.

The current forward modeling still has several limitations that require further improvement. First, we have not yet included the spatial distribution of the ground-level Hanle effect in Mercury's magnetosphere. In some regions, especially near the magnetotail current sheet, the magnetic field can be weak enough to enter the ground-level Hanle regime, where this effect may produce non-negligible changes in the spectral-line polarization. However, the present calculation retains only the \(q=0\) component of the ground-state density matrix and therefore cannot fully describe the ground-level Hanle effect. To treat these weak-field regions more accurately, future work should extend the current framework to solve the full ground-state density matrix, including all \(q\) components in the equations given in the Appendix, and then evaluate its influence on the spatial distribution of the polarization. Because of this limitation, the present forward modeling retains only regions where the Larmor precession rate exceeds the radiation pumping rate ($2\pi \nu_L>B_{lu}\bar J_0^0$). Second, in actual observations, solar radiation can be blocked by Mercury on the night side, reducing the local incident radiation and changing the optical-pumping conditions. Therefore, both the ground-level Hanle effect in weak-field regions and the occultation of the incident radiation by Mercury should be considered more carefully in future theoretical modeling and in the interpretation of observations. Despite these limitations, the present forward-modeling results demonstrate the strong potential of spectral-line polarimetric imaging for probing and constraining the large-scale magnetic-field structure of Mercury's magnetosphere.

The achievable spatial resolution is important for imaging Mercury's magnetosphere. Mercury's apparent angular diameter ranges from approximately $4.5^{\prime\prime}$ to $13^{\prime\prime}$ as seen from Earth, with a representative value of about $7^{\prime\prime}$. For an 80~\textrm{cm} ground-based telescope observing the Na~\textsc{i} $D_2$ line at 5891.6~\AA\ \citep{fei2021development}, the diffraction limit is approximately $0.19^{\prime\prime}$. For a $7^{\prime\prime}$ disk, this corresponds to about $0.054~R_{\rm M}$. With atmospheric seeing of approximately $1^{\prime\prime}$, the spatial scale is about $0.3~R_{\rm M}$, which is still sufficient to resolve large-scale magnetospheric structures. Moreover, THEMIS solar telescope obtained spatially resolved spectra of Mercury using a slit width of approximately $0.25^{\prime\prime}$ \citep{leblanc2009short,leblanc2013mercury}, demonstrating the feasibility of such observations. 

Besides the spatial resolution, another question is whether reliable Na~\textsc{i} D-line polarization images can be obtained within a reasonable exposure time. We therefore estimate the exposure time needed for Na~\textsc{i} D-line polarization. Following Equation (56) of \citet{trujillo2017physics}, the exposure time depends mainly on the Na~\textsc{i} D-line emission brightness, telescope aperture, total photon-detection efficiency, and spatial resolution. For estimating the exposure time required to detect a 1\% polarization signal at the \(3\sigma\) level, the numerator in Equation (56) of \citet{trujillo2017physics} is changed from \(10^7\) to \(10^5\), and $N_{\rm photons}\Delta \lambda$ in Equation (56) is replaced by $I_R10^6/4\pi$, with $I_R$ is the line-integrated emission intensity in Rayleighs. We first consider a 90~\textrm{cm} telescope with a total photon efficiency of 10\%. The Na~\textsc{i} D-line emission brightness decreases from about 500~kiloRayleighs near Mercury's surface to about 7~kiloRayleighs at $6~R_{\rm M}$ \citep{baumgardner2008imaging}. At a spatial resolution of $1^{\prime\prime}$, the required exposure time ranges from approximately 2.8 to 200~minutes from Mercury's surface to $6~R_{\rm M}$. Increasing the telescope diameter to 1.8~m reduces this range to approximately 42~seconds--50~minutes. For the 1.8-m telescope, spatially binning the observations to $2^{\prime\prime}$ further reduces the exposure time to approximately 10.5~seconds to 12.5~minutes. These results suggest that high spatial resolution can be achieved near Mercury, where the Na~\textsc{i} D-line emission is stronger. At larger distances, spatial binning can increase the photon counts and improve the temporal resolution. With this strategy, polarization images can be obtained with exposure times of approximately ten minutes or less. 

Regarding the observability of $I_{\rm D_2}/I_{\rm D_1}$, Figure~5 shows that variations at the level of about 1\% need to be resolved. The uncertainty in the intensity ratio depends on the photon statistics of both the D$_1$ and D$_2$ lines. For comparable photon counts in the two lines, achieving a 1\% precision at the $3\sigma$ level requires approximately twice the number of detected photons, and therefore roughly twice the exposure time, compared with a polarimetric measurement at the same precision. Moreover, the instrumental efficiency for intensity measurements may be higher if the polarization-analysis optics can be bypassed. Overall, the exposure time required to measure $I_{\rm D_2}/I_{\rm D_1}$ is expected to be comparable to that required for the polarization measurements. This suggests that the predicted variations in $I_{\rm D_2}/I_{\rm D_1}$ are potentially measurable.

For future observations, a reliable inversion of the magnetic-field orientation must account for the incident radiation, magnetic-field ambiguities, and LOS integration. First, the radiation-field tensors should be calculated from realistic solar Stokes spectra, including the intensity and polarization of the incident radiation. Realistic line profiles \citep{alsina2021solving} should also be considered, allowing for non-Maxwellian broadening and Doppler shifts of the line center \citep{del2018solar}. Second, in the GSA regime, the linear polarization can be either parallel or perpendicular to the projected magnetic field, resulting in a $90^\circ$ ambiguity. The projected magnetic-field direction also has an intrinsic $180^\circ$ ambiguity. These ambiguities may be reduced by combining the polarization measurements with a prior magnetic-field model and the solenoidal constraint $\nabla\cdot\boldsymbol{B}=0$. In addition, the observed Stokes signals are integrated along the LOS and contain contributions from regions with different sodium densities and magnetic-field orientations. A realistic exospheric model is therefore required to constrain the three-dimensional distribution of sodium. The inversion could use magnetic-field and exospheric models as initial constraints to identify physically consistent solutions.

\section{Conclusions}\label{sec5}

In this work, we use forward modeling to investigate whether spectral-line polarization can provide spatially resolved information on weak magnetic fields. Unlike in situ spacecraft measurements, spectropolarimetric imaging can provide two-dimensional information over an extended region. The achievable spatial and temporal resolutions are mainly determined by the telescope, instrument sensitivity, and exposure time.

We provide a framework for calculating the spectral-line polarization by solving the steady-state density-matrix equations, including anisotropic radiative pumping, magnetic precession, hyperfine structure, and collisions. The framework includes both GSA and the upper-level Hanle effect. We apply this framework to the Na~\textsc{i} D lines in Mercury's magnetosphere. Using three-dimensional distributions of sodium density and magnetic field, we generate synthetic images of the Na~\textsc{i} D-line intensity, linear polarization degree, polarization angle, and $I_{D_2}/I_{D_1}$ intensity ratio. At the magnetic-field strengths expected in Mercury's magnetosphere, the relevant regime is the GSA regime, where the polarization is mainly controlled by magnetic realignment of the ground state rather than by the upper-level Hanle effect. The Na~\textsc{i} D lines therefore primarily constrain the magnetic-field geometry rather than its strength. The synthetic images show that the polarization degree, polarization angle, and $I_{D_2}/I_{D_1}$ intensity ratio contain information on the global magnetic-field structure. The large-scale magnetospheric morphology remains identifiable at an angular resolution of approximately one arcsecond. These results demonstrate the potential of Na~\textsc{i} D-line polarimetry for imaging the global magnetic-field geometry of Mercury's magnetosphere. We also identify atomic and ionic lines that may be useful in other heliospheric environments. 

Recovering the three-dimensional magnetic field from future observations will require a dedicated inversion method that accounts for LOS integration, polarization ambiguities, realistic incident radiation, and the spatial distribution of the particle species. The inversion may be iteratively constrained and updated using prior magnetic-field and atmospheric models to obtain physically consistent solutions.

\begin{acknowledgments}
We thank the anonymous referee for the detailed comments and suggestions, which has helped improve the manuscript. We acknowledge the use of the multi-fluid MHD simulation data of Mercury’s magnetosphere provided by Y. Chen et al. (2024). We would like to acknowledge the use of ChatGPT to improve English grammar and sentence structure. The visualization interface of the web tool was developed with assistance from OpenAI Codex. C.H. is supported by the Alexander von Humboldt Foundation. S.Z. acknowledges the support from National Natural Science Foundation of China (NSFC) Excellent Young Scientists Fund (Overseas).
\end{acknowledgments}

\newpage

\appendix
\renewcommand{\theHequation}{\thesection.\arabic{equation}}

\section{Fine-Structure Levels with Collisions}\label{appendix_fine}
The statistical-equilibrium equations for the lower and upper levels are obtained by adding collisional terms to the equations of \citet{yan2006polarization}.
The population density of the upper level is governed by Equation (\ref{fine_u}), and the population density of the ground level is governed by Equation~(\ref{fine_l}). 
\begin{align}\label{fine_u}
\dot{\rho}^{k}_{q}(J_u) + 2\pi\mathrm{i}\nu_{\rm L}g_u q \rho^{k}_{q}(J_u)=
& - \sum_{J_l}A(J_u \rightarrow J_l)\rho^{k}_{q}(J_u) \\
& + \sum_{J_l k'}[J_l]\,\Bigg[\delta_{qq'}\delta_{k k'}\,p_{k'}(J_u,J_l)\,[B_{l u}\,\bar{J}^{0}_{0} + C^{(k)}]
\nonumber+ \sum_{Q q'}r_{k k'}(J_u,J_l,Q,q')\,B_{l u}\,\bar{J}^{2}_{Q}\Bigg]\,\rho^{k'}_{-q'}(J_l) .
\end{align}
\begin{align}\label{fine_l}
\dot{\rho}^{k}_{q}(J_l) + 2\pi\mathrm{i}\nu_{\rm L}g_l q \rho^{k}_{q}(J_l) =
& \sum_{J_u}p_k(J_u,J_l)\,[J_u]\,A(J_u \rightarrow J_l)\,\rho^{k}_{q}(J_u) \nonumber\\
& - \sum_{J_u k'}\Bigg[\delta_{k k'}\,[B_{l u}\,\bar{J}^{0}_{0} + C^{(0)}] \nonumber+ \sum_{Q q'}s_{k k'}(J_u,J_l,Q,q')\,B_{l u}\,\bar{J}^{2}_{Q} \Bigg]\,\rho^{k'}_{-q'}(J_l) \nonumber \\
& - \sum_{J_l'}C^{(0)}\rho^{k}_{q}(J_l)-D^{(k)}\rho^{k}_{q}(J_l) + \sum_{J_l'}p_k(J_l',J_l) [J_l'] C^{(k)} \rho^{k}_{q}(J_l').
\end{align}
The coefficients ($p_k$, $r_k$, $s_{kk'}$) are defined as:

$p_k(J_u,J_l) = (-1)^{J_u+J_l+1}\begin{Bmatrix}
J_l & J_l & k\\ 
J_u & J_u & 1
\end{Bmatrix}$,

$r_k(J_u,J_l, Q, q) = \sqrt{3[k,k',2]}
\begin{Bmatrix}
1 & J_u & J_l\\ 
1 & J_u & J_l\\
2 & k   & k' \\
\end{Bmatrix} \begin{pmatrix}
k & k' & K \\ 
q & q' & Q \\ 
\end{pmatrix}$,

$s_{kk'}(J_u,J_l, Q, q) = (-1)^{J_l-Ju+1}[J_l]\sqrt{3[k,k',K]}
\begin{pmatrix}
k & k' & 2 \\ 
q & q' & Q \\ 
\end{pmatrix}
\begin{Bmatrix}
1 & 1 & 2\\ 
J_l & J_l & J_u\\
\end{Bmatrix} \begin{Bmatrix}
k & k' & 2 \\ 
J_l & J_l & J_l \\ 
\end{Bmatrix}$,

In Equations (\ref{fine_u}) and (\ref{fine_l}), $A(J_u\rightarrow J_l)$ is the Einstein coefficient for spontaneous emission from the upper level ($J_u$) to the lower level ($J_l$). $B_{lu}$ is the excitation coefficient from the lower level to the upper level and can be calculated from $B_{lu} = \frac{[J_u]}{[J_l]}\frac{c^2}{2h\nu^3}A$, where $c$ is the speed of light in vacuum, $h$ is the Planck constant and $\nu$ is the frequency. 
$\bar{J}^K_Q$ denotes the radiation pumping term defined in Appendix (\ref{appendix_radiation}), while $\mathcal{J}^K_Q$ represents the outgoing-direction geometrical tensors to the observer, as described in Appendix (\ref{appendix_radiation}). The collisional terms are introduced in Appendix (\ref{appendix_collision}). The matrices with $\{ \}$ represents Wigner $6j$ and $9j$ symbols, the matrices with $()$ represents Wigner $3j$ symbol \citep{varshalovich1988quantum,yan2006polarization}, $[J] \equiv 2J+1$, and $[k,k',2] \equiv (2k+1)(2k'+1)(2\times2+1)$.

Combining Equations (\ref{fine_u}) and (\ref{fine_l}), we obtain a linear system for the population density of the ground levels. Here, we set $q = 0$ for the ground levels, since $\rho^{k}_{q \neq 0}(J_l) = 0$. Under the steady-state assumption, $\dot{\rho}^{k}_{q}(J_l) = \dot{\rho}^{k}_{q}(J_u) = 0$.

\begin{align}\label{fine_ul}
&\Bigg \{\sum_{J_uk'} \bigg \{p_k(J_u,J_l)\,\frac{[J_u]}{\sum_{J_{l}''}[A''(J_u\rightarrow J_l'') ]/[A(J_u\rightarrow J_l)]}\nonumber \\
& \times\sum_{J_l'}[J_l']\, \bigg[\delta_{k k'}\, p_{k'}(J_u,J_l')\, \,[B_{l u}\bar{J}^{0}_{0}+C^{(k)}(J_l'\rightarrow J_u)] + \sum_{Q} r_{k k'}(J_u,J_l',Q,0)\, \bar{J}^{2}_{Q} \bigg]\,\nonumber \\
& - \delta_{J_l J_l'}\bigg[ \delta_{k k'}\, [B_{l u}\bar{J}^{0}_{0}+C^{(0)}(J_l'\rightarrow J_u)]
\nonumber + \sum_{Q} s_{k k'}(J_u,J_l,Q,0)\, B_{l u}\, \bar{J}^{2}_{Q} \bigg]\,  \bigg \}  \nonumber \\
& - \sum_{J_l''}\delta_{J_l J_l'}\delta_{k k'}C^{(0)}(J_l'\rightarrow J_l'') - \delta_{J_l J_l'}\delta_{k k'}D^{(k)} + \sum_{J_l'}\delta_{k k'}p_{k'}(J_l',J_l) [J_l'] C^{(k)}(J_l'\rightarrow J_l) \Bigg \}\rho^{k'}_{0}(J_l') = 0.
\end{align}

After solving Equation (\ref{fine_ul}) and yielding $\rho^{k}_{0}(J_l)$, the absorption coefficient is then given by
\begin{align}\label{absorption_fine}
    \eta_i(\nu,\Omega) = \frac{h\nu_0}{4\pi}B_{lu}n\sqrt{[J_l]}\xi(\nu-\nu_0)\sum_{K}(-1)^K\omega^K_{J_lJ_u}\rho_0^K(J_l)\mathcal{J}_0^K(i,\Omega),
\end{align}
where $K = 0, 2$, $\omega^K_{J_lJ_u} \equiv \{1,1,K;J_l, J_l, J_u\}/\{1,1,0;J_l,J_l,J_u\}$, $i = 0, 1, 2$ corresponds to the Stokes parameters $I$, $Q$, and $U$, respectively. $\xi(\nu - \nu_0)$ denotes the line profile function of the spectral line at the center frequency $\nu_0$. n is the particle number density.

Combining Equations (\ref{fine_u}) and (\ref{fine_ul}) and yielding $\rho_q^k(J_u)$, the emission coefficient is given by:

\begin{equation}\label{appendix_emission_fine}
\epsilon_i(\nu, \Omega) = \frac{h \nu_0}{4\pi} A_{ul} n\sqrt{[J_u]}\,\xi(\nu - \nu_0)
\sum_{KQ} \omega^{K}_{J_u J_l}\, \rho^{K}_{Q}(J_u)\, \mathcal{J}^{K}_{Q}(i, \Omega),
\end{equation}
where $K = 0, 2$, and $Q=0,\pm1,\pm2$.

\subsection{Radiation Pumping Terms and Radiation Tensors}\label{appendix_radiation}

For a polarized point source \textcolor{black}{located in the direction $(\theta_r, \phi_r)$,  where $(\theta_r, \phi_r)$ specifies the direction from the particles toward the radiation source in the magnetic-field reference frame}, the radiation field \textcolor{black}{tensors} is \citep{landi2004polarization}:

\begin{equation}
\begin{aligned}
\bar{J}^0_0
&= I_*,
\\
\bar{J}^2_0
&= \frac{1}{2\sqrt{2}}
   \left(2-3\sin^2\theta_r\right)I_*
   -\frac{3}{2\sqrt{2}}\sin^2\theta_r\,\tilde{Q},
\\
\bar{J}^2_{\pm1}
&= \mp\frac{\sqrt{3}}{4}\sin 2\theta_r\,
   e^{\pm \mathrm{i}\phi_r}
   \left(I_*+\tilde{Q}
   \textcolor{black}{\mp}\frac{\mathrm{i}\tilde{U}}{\cos\theta_r}\right),
\\
\bar{J}^2_{\pm2}
&= \frac{\sqrt{3}}{4}\sin^2\theta_r\,
   e^{\pm 2\mathrm{i}\phi_r}I_*
   -\frac{\sqrt{3}}{4}
   \left(1+\cos^2\theta_r\right)
   e^{\pm 2\mathrm{i}\phi_r}\tilde{Q}
   \textcolor{black}{\pm}\frac{\sqrt{3}}{2}\mathrm{i}\cos\theta_r\,
   e^{\pm 2\mathrm{i}\phi_r}\tilde{U}.
\end{aligned}
\end{equation}
where $I_*$ is the incident specific intensity, which can be approximated by the Planck function,
\begin{equation}
I_* =
\frac{2h\nu^3}{c^2}
\left[
\exp\left(\frac{h\nu}{k_{\rm B}T}\right)-1
\right]^{-1},
\end{equation}
or derived from the observed solar irradiance. The Stokes parameters in the magnetic-field reference frame are
\begin{equation}
\tilde{Q}
=
\cos 2\gamma\,Q_*
-
\sin 2\gamma\,U_*,
\qquad
\tilde{U}
=
\cos 2\gamma\,U_*
+
\sin 2\gamma\,Q_*,
\end{equation}
where $Q_*$ and $U_*$ describe the linear polarization of the incident solar radiation, and $\gamma$ is the angle between the original polarization reference axis and the reference axis defined by the local magnetic field.

For an unpolarized point source from $(\theta_r, \phi_r)$, the above radiation field \textcolor{black}{tensors} with $Q_*
=U_*=0$ is reduced to :

\begin{equation}
\bar{J}^0_0 = I_*, \quad
\bar{J}^2_0 = \frac{1}{2\sqrt{2}} (2 - 3 \sin^2\theta_r) I_*,\quad \bar{J}^2_{\pm 1} = \mp \frac{\sqrt{3}}{4} \sin 2\theta_r \, e^{\pm \mathrm{i} \phi_r} I_*, \quad \bar{J}^2_{\pm 2} = \frac{\sqrt{3} }{4} \sin^2\theta_r \, e^{\pm \mathrm{i} 2\phi_r} I_*.
\end{equation}

The angles $\theta_r$ and $\phi_r$, which specify the direction of the incident radiation in the magnetic-field reference frame, can be determined from $\theta_0$, $\theta_B$, and $\phi_B$:

\begin{equation}
\begin{bmatrix}
\sin\theta_r\cos\phi_r\\
\sin\theta_r\sin\phi_r\\
\cos\theta_r\\
\end{bmatrix}
=
\begin{bmatrix}
\cos\theta_B \cos\phi_B & \cos\theta_B \sin\phi_B & -\sin\theta_B \\
-\sin\phi_B & \cos\phi_B & 0 \\
\sin\theta_B \cos\phi_B & \sin\theta_B \sin\phi_B & \cos\theta_B
\end{bmatrix}
\begin{bmatrix}
\sin\theta_0\\
0\\
\cos\theta_0\\
\end{bmatrix}
\end{equation}

The outgoing-direction geometrical tensors are given by:

\begin{align}
\mathcal{J}^0_0(i,\Omega) =
\begin{pmatrix}
1 \\
0 \\
0
\end{pmatrix},~ 
\mathcal{J}^2_0(i,\Omega) =
\frac{1}{\sqrt{2}}
\begin{pmatrix}
1 - 1.5 \sin^2\theta_B \\
-\frac{3}{2} \sin^2\theta_B  \\
0
\end{pmatrix},~
\mathcal{J}^2_{\pm 2}(i,\Omega) =
\sqrt{3} e^{\pm \mathrm{i} 2 \pi}
\begin{pmatrix}
\frac{\sin^2\theta_B }{4} \\
-\frac{1+\cos^2\theta_B }{4} \\
\mp \mathrm{i}\frac{\cos\theta_B }{2}
\end{pmatrix},~
\mathcal{J}^2_{\pm 1}(i,\Omega) =
\sqrt{3} e^{\pm \mathrm{i} \pi}
\begin{pmatrix}
\mp \frac{\sin 2\theta_B }{4} \\
\mp \frac{\sin 2\theta_B }{4} \\
- \mathrm{i} \frac{\sin\theta_B }{2}
\end{pmatrix}.
\end{align}

\subsection{Collisional Terms}\label{appendix_collision}
The collision rate $C^{(k)}$ with a unit of $\mathrm{s}^{-1}$ is calculated from \citep{zhang2021polarization}:

\begin{equation}
C^{(k)}(J_u\rightarrow J_l) = (-1)^{k} \frac{\begin{Bmatrix}
J_l & J_l & k \\
J_u & J_u & 1
\end{Bmatrix}}{\begin{Bmatrix}
J_l & J_l & 0 \\
J_u & J_u & 1
\end{Bmatrix}}\left(n_e\frac{8.629\times10^{-6}\Upsilon}{T^{1/2}[J_u]} + n_H\frac{1.1\times10^{-10}\Upsilon}{T^{1/2}[J_u]}\right),
\end{equation}

\begin{equation}
C^{(k)}(J_l\rightarrow J_u) = (-1)^{k} \frac{\begin{Bmatrix}
J_u & J_u & k \\
J_l & J_l & 1
\end{Bmatrix}}{\begin{Bmatrix}
J_u & J_u & 0 \\
J_l & J_l & 1
\end{Bmatrix}} \frac{[J_u]}{[J_l]}e^{\frac{E(J_l)-E(J_u)}{k_BT}} C^{(0)}(J_u\rightarrow J_l),
\end{equation}
where, $\Upsilon$ is the collisional strength and is of order unity \citep{pequignot1990populations,tayal1999collision,yan2006polarization}, $n_e$ is the electron number density with a unit of $\textrm{cm}^{-3}$. $n_H$ is the number density of Hydrogen with a unit of $\textrm{cm}^{-3}$, $E(J)$  denotes the energy at the level $J$, and the unit of temperature $T$ is kelvin. 

The depolarization rate due to collision is given by $D^{(k)} = 2\times 10^{-10} (\frac{13.6~\mathrm{eV}}{E_I-E(J)})^{0.8}[T(1+\frac{1}{\mu})]^{0.3}n_H$, where $E_I$ is the atomic ionization energy and $\mu$ is the atomic mass number.

\subsection{Upper-level Hanle Effect}

In Equation (\ref{fine_u}), the term $2\pi\mathrm{i}\nu_{\rm L} g_u q \rho^{k}_{q}(J_u)$ can be neglected when $2\pi \nu_{\rm L} g_u q \ll A$, corresponding to the GSA regime. When $2\pi \nu_{\rm L} g_u q \sim A$, this term must be retained, and the system enters the Hanle regime, in which magnetic precession modifies the population and coherence of the upper levels. The Landé factor is given by $g_u = 1 + \frac{J_u(J_u+1) + 3/4 - L(L+1)}{2J_u(J_u+1)}$ \citep{landi2004polarization}, where $L$ is the orbital angular momentum.

\section{Hyperfine Levels with Collisions}\label{appendix_hyper}

After taking into account the nuclear spin quantum number $I$, the total atomic angular momentum is given by $F=|J-I|,..,|J+I|$ with a step of 1. As in the fine-structure case, the statistical-equilibrium equations \citep{yan2007polarization} for the upper hyperfine levels, ($\rho_q^{k}(F_u, F_u')$), and the lower hyperfine levels, ($\rho_q^{k}(F_l)$), can be written by including the collisional terms.

\begin{align}\label{hyper_u}
\dot{\rho}^{k}_{q}(F_u,F_u') &+ 2\pi \mathrm{i} \nu_L \Gamma_{FF'} q \rho^{k}_{q}(F_u,F_u') +{2\pi\mathrm{i}\nu_{F_uF_u'}}\rho^{k}_{q}(F_u,F_u') =\nonumber\\
&-(A(J_u \rightarrow J_l)+C^{(0)}(F_uF_u' \rightarrow F_l)+D^{(k)})\rho^{k}_{q}(F_u,F_u') \nonumber\\
&+ [J_l]\sum_{F_l' k' q'}\left(\delta_{-qq'}\delta_{kk'} p_{k'} (B_{lu}\bar{J}^{0}_{0}+C^{(k)}(F_l'\rightarrow F_uF_u' ))+ \sum_{Q}r_{kk'} B_{lu}\bar{J}^{2}_{Q}\right)\rho^{k'}_{-q'}(F_l')\nonumber\\
& - \sum_{F_u''}\frac{1}{2}(C^{(0)}(F_u\rightarrow F_u'')+C^{(0)}(F_u'\rightarrow F_u''))\rho^{k}_{q}(F_u,F_u') \nonumber\\
& +\sum_{F_u''} \sum_{F_u'''} \frac{1}{2}\{(p_k(F_u'',F_u) [J_u] C^{(k)}(F_u''\rightarrow F_u) \rho^{k}_{q}(F_u'',F_u''') + (p_k(F_u'',F_u') [J_u] C^{(k)}(F_u''\rightarrow F_u') \rho^{k}_{q}(F_u'',F_u''')\},
\end{align}

\begin{align}\label{hyper_l}
\dot{\rho}^{k}_{q}(F_l) + 2\pi\mathrm{i}\nu_L g_l q \rho^{k}_{q}(F_l) &=
\sum_{J_u,F_u,F_u'} p_k [J_u] (A(J_u \rightarrow J_l)+C_S^{(k)}(F_uF_u'\rightarrow F_l))\rho^{k}_{q}(F_u,F_u') \nonumber\\
&- \sum_{J_u,F_u,k'} \left(\delta_{kk'} (B_{lu}\bar{J}^{0}_{0}+C^{(0)}(F_l\rightarrow F_u)) + s_{kk'} B_{lu}\bar{J}^{2}_{0}\right)\rho^{k'}_{-q'}(F_l) \nonumber\\
& - \sum_{F_l'}C^{(0)}(F_l\rightarrow F_l')\rho^{k}_{q}(F_l)-D^{(k)}\rho^{k}_{q}(F_l) + \sum_{F_l'}p_k(F_l',F_l) [J_l'] C^{(k)}(F_l'\rightarrow F_l) \rho^{k}_{q}(F_l').
\end{align}

$p_k = [F_l](-1)^{F_u'+F_l+k+1}
\begin{Bmatrix} 
F_l & F_l & k \\
F_u & F_u' & 1
\end{Bmatrix} \sqrt{[F_u,F_u']}
\begin{Bmatrix}
J_u & J_l & 1 \\
F_l & F_u & I
\end{Bmatrix}
\begin{Bmatrix}
J_u & J_l & 1 \\
F_l & F_u' & I
\end{Bmatrix}
$,

$r_{kk'} = (-1)^{k'+q'}\sqrt{3[k,k',2]}\sqrt{[F_u,F_u']}[F_l']
\begin{pmatrix}
k & k' & 2 \\
q & q' & Q
\end{pmatrix}
\begin{Bmatrix}
1 & F_u & F_l' \\
1 & F_u'& F_l' \\
2 & k   & k'
\end{Bmatrix}
\begin{Bmatrix}
J_u & J_l & 1 \\
F_l' &F_u & I \\
\end{Bmatrix}
\begin{Bmatrix}
J_u & J_l & 1 \\
F_l' &F_u' & I \\
\end{Bmatrix}
$,

$
s_{kk'} = (-1)^{J_u-I+F_l+q'+1}[J_l,F_l]\sqrt{3[k,k',2]}
\begin{pmatrix}
k & k' & 2 \\
q & q' & Q
\end{pmatrix}
\begin{Bmatrix}
J_l & J_l & 2 \\
1 &1 & J_u \\
\end{Bmatrix}
\begin{Bmatrix}
J_l & J_l & 2 \\
F_l & F_l & I \\
\end{Bmatrix}
\begin{Bmatrix}
k & k' & 2 \\
F_l & F_l & F_l \\
\end{Bmatrix}
\frac{1}{2}[1+(-1)^{k+k'+2}]
$.

In which $\nu_{F_uF_u'}$ is the energy difference between $E(F_u)$ and $E(F_u')$, the radiation pumping terms and radiation tensors are the same as those for a fine-structure atom (see Appendix (\ref{appendix_radiation})). The collisional terms ($C^{(0)}$, $C^{(k)}$, and $C_{S}^{(k)}$) are described in Appendix (\ref{appendix_collision}).

Combining Equations (\ref{hyper_u}) and (\ref{hyper_l}) and $\dot{\rho}^{k}_{q}(F_u,F_u') = \dot{\rho}^{k}_{q}(F_l)= 0$, we obtain the population equations for upper levels and ground levels:

\begin{align}\label{hyper_us}
\rho^{k}_{q}(F_u,F_u') &= \frac{1}{M+2\pi\mathrm{i}\nu_L \Gamma_{FF'} q } [J_l]\sum_{F_l' k'}\left(\delta_{q0}\delta_{kk'} p_{k'} (B_{lu}\bar{J}^{0}_{0}+C^{(k)}(F_l'\rightarrow F_uF_u'))+ \sum_{Q}r_{kk'} B_{lu}\bar{J}^{2}_{Q}\right)\rho^{k'}_{0}(F_l') +\frac{N}{M+2\pi\mathrm{i}\nu_L \Gamma_{FF'} q }
\end{align}

\begin{align}\label{hyper_ul}
& \Bigg \{\sum_{J_u,F_u,F_u'} p_k [J_u] (A +C^{(k)}(F_uF_u'\rightarrow F_l))\bigg \{ \frac{1}{M} [J_l]\sum_{F_l' k'}\bigg ( \delta_{kk'} p_{k'} (B_{lu}\bar{J}^{0}_{0}+C^{(k)}(F_l'\rightarrow F_uF_u'))+ \sum_{Q}r_{kk'} B_{lu}\bar{J}^{2}_{Q}\bigg )+\frac{N}{M}\bigg \} \nonumber\\
& - \sum_{J_u,F_u,k'} \delta_{F_lF_l'}\bigg(\delta_{kk'}(B_{lu}\bar{J}^{0}_{0}+C^{(0)}(F_l\rightarrow F_u)) + \sum_{Q}s_{kk'} B_{lu}\bar{J}^{2}_{Q}\bigg) \nonumber\\
& - \sum_{F_l''}\delta_{F_lF_l'}\delta_{kk'}C^{(0)} (F_l\rightarrow F_l')-\delta_{F_lF_l'}\delta_{kk'}D^{(k')} + \sum_{F_l'}\delta_{kk'}p_k(F_l',F_l) [J_l'] C^{(k)}(F_l'\rightarrow F_l) \Bigg \} \rho^{k'}_{0}(F_l') =0,
\end{align}

where 
\begin{align}
M = &2\pi\mathrm{i}\nu_{F_uF_u'}+A \nonumber +C^{(0)}(F_uF_u' \rightarrow F_l)+D^{(k)} + \sum_{F_u''}\frac{1}{2}\bigg(C^{(0)}(F_u\rightarrow F_u'')+C^{(0)}(F_u'\rightarrow F_u'')\bigg), \nonumber\\
N= &\sum_{F_u''}\sum_{F_u'''} \bigg(\frac{1}{2}p_k(F_u'',F_u) [J_u] C^{(k)}(F_u''\rightarrow F_u) + \frac{1}{2} p_k(F_u'', F_u') [J_u] C^{(k)}(F_u''\rightarrow F_u')\bigg )\rho_q^k(F_u'',F_u''').\nonumber
\end{align}

Because $N$ contains the upper-level density elements, Equation (\ref{hyper_ul}) is not formally closed and could be solved iteratively. In the first step, we set $N=0$ and solve Equations (\ref{hyper_us}) and (\ref{hyper_ul}) to obtain $\rho_q^k(F_u'',F_u''')$. We then use these solutions as known input parameters in the next iteration of Equations (\ref{hyper_us}) and (\ref{hyper_ul}), and repeat the procedure until convergence. The iterative calculation is considered converged when the maximum relative change in the lower-level density-matrix components, all upper-level density-matrix components, and the collisional terms of the upper levels becomes smaller than a prescribed tolerance ($\sim 10^{-8}$). For the online tool, user can specify both the convergence tolerance and the maximum number of iterations. In the forward modeling presented in Section~\ref{sec3}, all collisional terms are neglected.

After solving Equation (\ref{hyper_ul}) and yielding $\rho_0^k(F_l)$, the absorption coefficient is then given by:

\begin{equation}\label{absorption_hyper}
\eta_i(\nu,\Omega)
=
\frac{h\nu_0}{4\pi}
B n \xi(\nu-\nu_0)[J_l]
\sum_{KF_l}
[F_l]\sqrt{3}(-1)^{1-J_u+I+F_l}
\begin{Bmatrix}
J_l & J_l & K \\
F_l & F_l & I
\end{Bmatrix}
\begin{Bmatrix}
1 & 1 & K \\
J_l & J_l & J_u
\end{Bmatrix}
\rho^K_0(F_l)
\mathcal{J}^K_0(i,\Omega),
\end{equation}
where $K = 0, 2$, $i = 0, 1, 2$ corresponds to the Stokes parameters $I$, $Q$, and $U$, respectively, and $\xi(\nu - \nu_0)$ denotes the line profile function of the spectral line at the center frequency $\nu_0$.

Combining Equations (\ref{hyper_us}) and (\ref{hyper_ul}) and yielding $\rho_q^k(F_u,F_u')$, the emission coefficient is given by:

\begin{equation}\label{appendix_emission_hyper}
\epsilon_i(\nu,\Omega)
=
\frac{h\nu_0}{4\pi}
A n \xi(\nu-\nu_0)[J_u]
\sum_{KQF_uF'_uF_l}
[F_l]\sqrt{3[F_u,F'_u]}(-1)^{F_u+F_l+1}
\begin{Bmatrix}
J_u & J_l & 1 \\
F_l & F_u & I
\end{Bmatrix}
\begin{Bmatrix}
J_u & J_l & 1 \\
F_l & F'_u & I
\end{Bmatrix}
\begin{Bmatrix}
F_u & F'_u & K \\
1 & 1 & F_l
\end{Bmatrix}
\rho^K_Q(F_u,F'_u)
\mathcal{J}^K_Q(i,\Omega),
\end{equation}
where $K = 0, 2$, $Q=0,\pm1,\pm2$, and $\mathcal{J}^K_Q(i,\Omega)$ is the outgoing-direction geometrical tensors described in Appendix~\ref{appendix_radiation}.

\subsection{Upper-level Hanle Effect}

In Equation (\ref{hyper_us}), the term $2\pi\mathrm{i}\nu_L \Gamma_{FF'} q$ can be neglected when $2\pi\mathrm{i}\nu_L \Gamma_{FF'} q \ll A$, corresponding to the pure GSA regime. When $2\pi\mathrm{i}\nu_L \Gamma_{FF'} q \sim A$, this term must be retained, and the system enters the upper Hanle effect regime, in which magnetic precession modifies the population and coherence of the upper levels. The Land\'e factor is given by $\Gamma_{FF'} = (-1)^{1+J_u+I+F_u} g_{u} \sqrt{J_u(J_u+1)(2J_u+1)(2F_u+1)(2F_u'+1)}\begin{Bmatrix}
F_u & F_u' & 1 \\
J_u & J_u & I \\
\end{Bmatrix}$ \citep{landi2004polarization}.

\section{Rotation of the Reference Axis for the Stokes Parameters}\label{appendix_rot}

The reference axis for the Stokes parameters calculated from Appendices \ref{appendix_fine} and \ref{appendix_hyper} is chosen to be along the magnetic field. To rotate the Stokes parameters counterclockwise by an angle $\gamma$ to an instrument reference axis in the plane of the sky, we apply

\begin{equation}\label{eq_rot_QU}
\begin{bmatrix}
I_{\rm instrument}\\
Q_{\rm instrument}\\
U_{\rm instrument}\\
\end{bmatrix}
=
\begin{bmatrix}
1 &            0 &           0 \\
0 &  \cos2\gamma & \sin2\gamma \\
0 & -\sin2\gamma & \cos2\gamma \\
\end{bmatrix}
\begin{bmatrix}
I_{\rm theory}\\
Q_{\rm theory}\\
U_{\rm theory}\\
\end{bmatrix}
\end{equation}

For the forward modeling of polarization of Mercury's magnetosphere, we adopt a fixed $z$-axis as the observational reference axis. Because the magnetic field direction varies across space, Equation (\ref{eq_rot_QU}) is applied at each spatial location to convert the calculated Stokes parameters from the local magnetic field frame to the fixed observational frame, and $\gamma$ is chosen as the angle between the local magnetic field and the z-axis in the plane of the sky.


\end{document}